\documentclass[a4paper,11pt]{article}
\usepackage{amsmath}
\usepackage{amsfonts}
\usepackage{amssymb}
\usepackage{bm}
\usepackage[utf8x]{inputenc}
\usepackage{ae}
\usepackage[T1]{fontenc}
\usepackage[english]{babel}
\usepackage{graphicx}

\usepackage{makecell}
\usepackage{placeins}
\usepackage{tabularx}
\usepackage{tabulary}
\usepackage{setspace}
\usepackage{tablefootnote}
\usepackage[flushmargin]{footmisc}
\usepackage[comma]{natbib}
\usepackage{array}
\usepackage{float}
\usepackage{caption}
\usepackage{tikz}
\usetikzlibrary{arrows.meta, positioning, shapes.multipart}
\usepackage[a4paper,left=2.5cm,right=2.5cm,top=3cm,bottom=3cm]{geometry}
\usepackage[colorinlistoftodos]{todonotes}
\usepackage{pdfpages}
\usepackage{longtable}
\usepackage{adjustbox}
\usepackage{booktabs}
\usepackage{multirow}
\usepackage{lscape}
\usepackage[colorlinks=true, allcolors=blue]{hyperref}

\usepackage[flushleft]{threeparttable}
\usepackage{inputenc}



\begin{document} \doublespacing \pagestyle{plain}
	
	\def\ci{\perp\!\!\!\perp}
	\begin{center}
		
		{\LARGE Visible or Covert? The Causal Effect of Inspector Visibility on Fare Evasion Detection: A Causal Machine Learning and Policy Learning Approach}

		{\large \vspace{0.4cm}}
		
		{\large Hannes Wallimann$^*$, Cédric Brütsch$^*$$^{\dagger}$ and Martin Huber$^{\dagger}$}\medskip

		{\small {$^*$University of Applied Sciences and Arts Lucerne, Competence Center for Mobility} \medskip }
		
		{\small {$^{\dagger}$University of Fribourg, Dept. of Economics} \medskip }
		
		{\large \vspace{0.3cm}}
		
		{\large 2026}\medskip

	\end{center}
	
	\smallskip

	\noindent \textbf{Abstract:} {\small
		Fare evasion generates substantial revenue losses for public transport operators and is typically combated through fare inspections, yet little is known about how the mode of inspection---uniformed versus plainclothes---affects detection efficiency. Using a unique dataset of 178,686 individual inspection records from PostAuto, the largest regional bus operator in Switzerland, we apply causal machine learning to estimate the causal effect of inspector visibility on inspection efficiency, defined as detected fare evaders per inspection hour. Our results indicate that wearing a uniform reduces efficiency by 0.156 incidents per hour, or approximately 26\% of the sample average.  This can also be interpreted as the immediate deterrent effect of visibility: passengers who would otherwise be detected may instead choose to comply. Heterogeneity analyses find that this effect is larger in areas with a high share of travelcard holders. When applying optimal policy learning to optimally target subgroups by one or the other treatment, plainclothes inspections are recommended for the large majority of contexts (91.8\%). Uniformed inspections are only suggested for lines with a low share of foreign residents but a relatively large population, and for lines with a high share of foreign residents and many total inspection hours in the previous month.		
	}
	
	{\small \smallskip }
	{\small \smallskip }
	{\small \smallskip }
	
	
	{\small \smallskip }
	{\small \smallskip }
	{\small \smallskip }
	
	{\footnotesize \noindent \textbf{Acknowledgments:} We are indebted to Bruno Zwyssig, Timon Heiland, Luigi Orlando, Simon Schüpbach, Reto Steiner, and Benedetto Barabino for their helpful comments. We used large language models (mainly Claude by Anthropic) as supportive tools for programming assistance and linguistic refinement of the manuscript. All outputs were carefully reviewed and verified by the authors, who take responsibility for the content presented.}
	
	\bigskip
	\bigskip
	\bigskip
	\bigskip
	
	{\small {\scriptsize 
\begin{spacing}{1.5}\noindent  
\textbf{Addresses for correspondence:} Hannes Wallimann, \href{mailto:hannes.wallimann@hslu.ch}{hannes.wallimann@hslu.ch}; Cédric Brütsch, \href{mailto:cedric.bruetsch@hslu.ch}{cedric.bruetsch@hslu.ch}; Martin Huber, \href{mailto:martin.huber@unifr.ch}{martin.huber@unifr.ch}.
\end{spacing}
			
		}\thispagestyle{empty}\pagebreak  }

	{\small \renewcommand{\thefootnote}{\arabic{footnote}} %
		\setcounter{footnote}{0}  \pagebreak \setcounter{footnote}{0} \pagebreak %
		\setcounter{page}{1} }
	
\section{Introduction}\label{introduction}

Public transport companies can adopt different ticketing systems. One common approach is the proof-of-payment system, an open-access system that allows passengers to board transport vehicles without prior ticket inspection, with fares verified through random spot checks by inspectors. This non-excludability gives rise to fare evasion, defined as the deliberate act of travelling on public transport without purchasing, validating, or correctly selecting the required ticket \citep[see, e.g.,][]{barabino2020fare}.

Fare evasion can generate substantial economic losses for public transport companies and may also damage their corporate image. To enforce fare payment, public transport companies typically impose monetary fines on detected fare evaders and organise inspection teams with respect to timing and location \citep{becker1968crime, barabino2020fare, buehler2017payment, barabino2014fare}. However, fare inspection strategies are not only defined by their intensity and location \citep[see, e.g., ][]{killias2009effects,barabino2014fare,barabino2019moving} but also by their mode of implementation, particularly whether inspections are conducted by uniformed (visible) or plainclothes (covert) staff \citep{barabino2020fare}. Visible enforcement is expected to increase the perceived probability of inspection, thereby discouraging opportunistic fare evasion at the 'deterrence margin', that is, before an inspection ever takes place \citep{becker1968crime}. Yet, this visibility may simultaneously enable strategic adaptation by passengers, who can evade or adjust their behaviour when enforcement is observable. In contrast, plainclothes inspections are less salient but potentially more effective at the 'detection margin', that is, at identifying actual offenders once an inspection is underway, as they limit opportunities for behavioural adjustment. Empirical evidence in public transport remains scarce, but existing studies suggest that inspection outcomes depend on enforcement visibility: using data from Lyon, \citet{egu2020comparable} show that inspection records are inherently biased by enforcement practices, as visible controls may underestimate true evasion due to systematic avoidance behaviour. Similarly, \citet{keuchel2020effects}, to the best of our knowledge the only study directly comparing uniformed and plainclothes inspections, provide indicative evidence that inspection strategies influence observed evasion patterns, without isolating the causal impact of inspector visibility, as their identification relies on a single company-wide switch in inspector clothing at one point in time. In contrast, our study features substantial cross-sectional variation in inspection mode across many lines, months, and time slots within the same operator. 

To address this gap, we analyse a unique inspection dataset from PostAuto, the largest operator of regional bus services in Switzerland. We complement this dataset with potential determinants of fare evasion, including the share of the population holding a public transport subscription, the share of foreign residents, the youth dependency ratio, the number of inhabitants, and the social assistance rate. Our analysis is conditional on inspections being conducted and therefore does not generalise to contexts in which no inspection activity takes place. We identify the causal effect of inspection strategy under a selection-on-observables assumption, arguing that---conditional on the determinants of fare evasion and prior inspection activity---the assignment of inspection strategy can be regarded as good as random.

In a first step, we apply causal machine learning \citep[i.e., the causal forest, see][]{athey2019generalized} to estimate the effect of the binary treatment of uniformed (\textit{Präsenzkontrolle}) versus plainclothes (\textit{Normalkontrolle}) inspections on inspection efficiency, defined as the number of detected fare evaders per unit of inspection time. This approach is particularly well suited to our setting: our dataset contains a comparatively large set of line-level contextual characteristics, and it is not obvious \textit{a priori} which subset of these characteristics is required to satisfy the selection-on-observables assumption. Selecting this subset via ad hoc rules would expose our estimates to a pre-testing (or ``snooping'') problem, whereas the causal forest learns the relevant characteristics from the data without compromising the validity of the resulting estimates. In other words, we estimate the efficiency losses at the detection margin associated with deploying uniformed rather than plainclothes inspectors. 

We further examine treatment effect heterogeneity across all observed characteristics. Specifically, we estimate the Best Linear Predictor (BLP) of the conditional average treatment effect (CATE), which tests whether statistically significant heterogeneity exists by projecting the CATE linearly onto a pre-selected set of contextual characteristics. In addition, we estimate the Sorted Group Average Treatment Effects (GATES), which quantify the average treatment effect across groups ranked by their CATE and subsequently test for significant differences between groups \citep[see][]{Chernozhukov2025}. 


In a second step, we apply optimal policy learning \citep[][]{kitagawa2018should,athey2021policy} to determine the inspection strategy that maximises inspection efficiency across contexts. Optimal policy learning aims to optimally allocate an intervention across subgroups based on the size of their CATEs. We implement this using a policy tree, which learns both the optimal subgroup segmentation and the optimal intervention assignment in a data-driven way. 

Our results suggest that plainclothes inspections are on average significantly more effective at the detection margin than uniformed inspections. We estimate an average treatment effect of wearing a uniform of $-$0.156 incidents per hour (i.e., a relative reduction of approximately 26\% compared to the sample mean). While the distribution of estimated CATEs reveals some variation across contexts, the effect is consistently negative across all observed characteristics: Sorted Group Average Treatment Effects provide no evidence that uniformed inspections outperform plainclothes inspections in any of the five subgroups. These findings suggest that the higher detection efficiency of plainclothes inspections is a robust and pervasive phenomenon across the PostAuto network.

Because the analysis applying causal machine learning reveals  evidence of systematic treatment effect heterogeneity, we apply optimal policy learning to examine whether a data-driven allocation rule can identify contexts in which uniformed inspections are preferable. Applying a policy tree \citep{athey2021policy}, we find that plainclothes inspections are recommended for the large majority of contexts (91.8\%). Uniformed inspections are only suggested in two settings: for lines with a low share of foreign residents but a relatively large population, and for lines with a high share of foreign residents and many total inspection hours in the previous month.

While machine learning methods have recently been applied to fare evasion research---for instance, to classify fare evader profiles \citep{barabino2025comparing}, to compare econometric and machine learning approaches for predicting fare evasion frequency \citep{barabino2025comparingfareevasion}, to predict fare evasion risk in real time \citep{barabino2024toward}, and to optimise inspector routing \citet{delfau2018optimization}---these approaches remain non-causal in nature. A further contribution of our study is, to the best of our knowledge, the first application of causal machine learning and optimal policy learning to the study of fare inspection strategies.

The paper proceeds as follows. In Section \ref{background}, we discuss the case of Switzerland. Section \ref{Litrev} presents the relevant literature including the determinants of fare evasion. In Section \ref{Data}, we describe our unique dataset. Section \ref{Identification_estimation} outlines the identification and estimation of the effects as well as a detailed discussion of the behavioural assumptions. Section \ref{Results} presents the results, which we discuss in Section \ref{Discussion}. 

\section{Institutional Background}\label{background}

In 2025, public transport in Switzerland generated total revenues of CHF 7.04 billion. In total, transport companies sold approximately 295 million tickets, with ticket sales accounting for 31\% of total revenue. Single tickets accounted for the largest share of total sales volume, representing around 71\% of all tickets sold. Overall, 76.4\% of tickets were sold through digital channels, with approximately one quarter of these transactions conducted via automatic-ticketing solutions, where passengers use their devices to check in before boarding and check out at the end of the trip. Offline sales channels, such as ticket machines and ticket counters, have lost importance in recent years.\footnote{All figures in this paragraph are based on \url{https://www.allianceswisspass.ch/de/asp/News/Newsmeldung?filterCategory=4-20&newsid=997}, accessed on February 25, 2026.}

Single tickets also account for the largest share of revenues, representing approximately 31\%. They are followed by the so-called General Abonnement (GA), an unlimited-use subscription that allows travel for one month or one year across almost the entire Swiss public transport network, accounting for nearly 20\% of total revenues. At the end of 2024, around 425,000 individuals held a GA in Switzerland, corresponding to approximately 4.7\% of the resident population.\footnote{See \url{https://reporting.sbb.ch/en/finance}, accessed on February 25, 2026.} In terms of revenues, annual subscriptions offered by regional transport associations---i.e., unlimited-use passes valid within designated regions or tariff zones---rank next, followed by day passes. In addition, approximately 3.45 million individuals (about 38\% of the Swiss population) hold a Half Fare Travelcard (Halbtax), granting a 50\% price reduction on single tickets for both travel within regional tariff associations and between them.\footnote{See \url{https://reporting.sbb.ch/en/finance}, accessed on February 25, 2026.} Other popular products include Halbtax PLUS and the GA Night. A total of about 190,000 Halbtax PLUS packages and 123,000 GA Night subscriptions were in circulation. The former provides prepaid public transport credits that can be used to purchase a range of tickets, excluding season tickets. The latter costs CHF 99 and allows individuals under the age of 25 to travel free of charge from 7 p.m. onwards across almost the entire Swiss public transport network.\footnote{All figures in this paragraph, unless otherwise stated, are based on \url{https://www.allianceswisspass.ch/de/asp/News/Newsmeldung?filterCategory=4-20&newsid=997}, accessed on February 25, 2026.}

Despite the widespread availability of these products, fare evasion remains a significant challenge. According to estimates by Alliance SwissPass (the association of public transport companies in Switzerland), fare evasion resulted in annual revenue losses of up to CHF 200 million for the Swiss public transport system in 2024. According to the industry association, this figure is expected to remain largely unchanged in 2025.\footnote{See \url{https://www.srf.ch/news/schweiz/ohne-billett-unterwegs-starker-anstieg-bei-den-erwischten-schwarzfahrern}, accessed on February 25, 2026.} Therefore, this figure corresponds to approximately 2.8\% of total revenues.

In 2025, a total of 1{,}173{,}295 passengers in Switzerland were detected travelling without a valid or fully valid ticket.\footnote{See \href{https://www.srf.ch/news/schweiz/ohne-billett-unterwegs-starker-anstieg-bei-den-erwischten-schwarzfahrern}{SRF News}, accessed on February 25, 2026.} This number has increased annually since 2019. Since that year, passengers travelling without a valid ticket have been centrally recorded by all Swiss public transport operators through the central information system SynServ.\footnote{See \href{https://www.voev.ch/de/Service/Ombudsstelle-oV/Informationen-zu-SynServ}{VöV}, accessed on February 25, 2026.}

SynServ is operated by PostAuto on behalf of the national tariff association Alliance SwissPass. Passengers repeatedly travelling without a valid ticket are subject to progressively higher fines. For a first offence, the surcharge, including a flat-rate fare supplement, amounts to CHF 100; for a second offence, CHF 140; and for a third offence, CHF 170. The collection of these surcharges, as well as any discretionary reductions (e.g., goodwill adjustments), is the responsibility of the individual transport companies.

PostAuto is also the largest operator of regional bus services in Switzerland. In 2025, the company operated 942 lines with a network length of approximately 18{,}000 km, serving around 11{,}500 stops. PostAuto generated an operating income of CHF 1{,}168 million and transported 189.2 million passengers in Switzerland.\footnote{For all figures, see \url{https://www.postauto.ch/en/about-us-and-news/organization/facts-and-figures}, accessed on May 18, 2026.} 

In general, three levels of fare inspection can be distinguished \citep{zvv_t651_8_2025}. At the first level, the primary objective is to check as many passengers as possible. At the second level, the inspection is not recognisable as such prior to its announcement: the appearance and behaviour of inspectors give no indication that a fare check is taking place. At the third level, passengers cannot evade the inspection: once the fare check has been announced or inspectors have become visible, evasion is no longer possible. Refusal is not an option, ideally enforced through a police presence, and all alighting passengers are subject to inspection. PostAuto's inspections are predominantly conducted at the second level, which forms the institutional basis for the plainclothes strategy examined in this study.

\section{Literature review}\label{Litrev}
Fare inspections combined with the imposition of fines on passengers caught without a valid ticket constitute the most widely adopted strategy to combat fare evasion in proof-of-payment transit systems worldwide \citep{barabino2024fare}. As our study examines how the choice of inspection strategy (i.e., plainclothes versus uniformed staff) affects inspection efficiency, and which contextual factors moderate this effect, we review the literature along two dimensions: the determinants of fare evasion frequency, which inform the contextual characteristics included in our analysis, and existing evidence on the effectiveness of inspection strategies, which motivates our research question.

\subsection{Determinants of Fare Evasion Frequency}\label{Lit_determinants}

In the following, we present studies investigating the attributes of potential fare evaders. These attributes are also summarised in Table \ref{tab:fare_evasion_attributes}. 

\begin{table}[htbp]
	\tiny
	\centering
	\caption{Attributes Associated with Fare Evasion}
	\label{tab:fare_evasion_attributes}
	\begin{tabular}{lll}
		\toprule
		\textbf{Attribute} & \textbf{Source} & \textbf{Context} \\
		\midrule
		
		\multicolumn{3}{l}{\textit{Socio-demographics}} \\
		\addlinespace
		
		\multicolumn{3}{l}{\textbf{Age} \tiny(younger passengers are more likely to evade fares)} \\
		& \citet{bucciol2013unethical} & Bus passengers, Reggio Emilia (Italy) \\
		& \citet{barabino2015determinants} & Italian public transport company \\
		& \citet{cools2018identification} & Flanders (Belgium), survey data \\
		& \citet{cantillo2022fare} & Santiago (Chile), bus vs.\ metro \\
		& \citet{barabino2022assessing} & Cagliari (Italy) \\
		& \citet{barabino2023evaluating} & Italy, bus company \\
		\addlinespace
		
		\multicolumn{3}{l}{\textbf{Gender} \tiny(males more likely to evade fares)} \\
		& \citet{bucciol2013unethical} & Bus passengers, Reggio Emilia (Italy) \\
		& \citet{barabino2015determinants} & Italian public transport company \\
		& \citet{cools2018identification} & Flanders (Belgium), survey data \\
		& \citet{cantillo2022fare} & Santiago (Chile) \\
		& \citet{barabino2022assessing} & Cagliari (Italy) \\
		& \citet{barabino2023evaluating} & Italy, bus company \\
		\addlinespace
		
		\addlinespace
		\multicolumn{3}{l}{\textbf{Unemployed} \tiny(unemployed passengers more likely to evade fares)} \\
		& \citet{bucciol2013unethical} & Bus passengers, Reggio Emilia (Italy) \\
		& \citet{barabino2015determinants} & Italian public transport company \\
		
		\addlinespace
		\multicolumn{3}{l}{\textbf{Student} \tiny(students more likely to evade fares)} \\
		& \citet{barabino2015determinants} & Italian public transport company \\
		
		\addlinespace
		\multicolumn{3}{l}{\textbf{Low education} \tiny(lower education associated with more fare evasion)} \\
		& \citet{barabino2015determinants} & Italian public transport company \\
		& \citet{barabino2023evaluating} & Italy\\
		
		\addlinespace
		\multicolumn{3}{l}{\textbf{Immigrant} \tiny(non-European immigrants more likely to evade fares)} \\
		& \citet{bucciol2013unethical} & Reggio Emilia (Italy) \\
		
		\midrule
		\multicolumn{3}{l}{\textit{Travel behaviour and trip characteristics}} \\
		\addlinespace
		
		\multicolumn{3}{l}{\textbf{Short trips} \tiny(shorter trips associated with higher evasion)} \\
		& \citet{bucciol2013unethical} & Reggio Emilia (Italy) \\
		
		\addlinespace
		\multicolumn{3}{l}{\textbf{Occasional passenger} \tiny(occasional users more likely to evade fares)} \\
		& \citet{bucciol2013unethical} & Reggio Emilia (Italy) \\
		& \citet{barabino2023evaluating} & Italy\\
		
		\midrule
		\multicolumn{3}{l}{\textit{Service, system, and operational context}} \\
		\addlinespace
		
		\multicolumn{3}{l}{\textbf{Route characteristics} \tiny(evasion likelihood varies by route)} \\
		& \citet{lee2011uncovering} & San Francisco (USA), railway transit \\
		
		\addlinespace
		\multicolumn{3}{l}{\textbf{Time of day} \tiny(higher evasion during evening/night periods)} \\
		& \citet{lee2011uncovering} & San Francisco (USA) \\
		& \citet{cantillo2022fare} & Santiago (Chile) \\
		
		\addlinespace
		\multicolumn{3}{l}{\textbf{Vehicle characteristics (transport mode)} \tiny(evasion differs between bus and metro; higher in longer, multi-door vehicles)} \\
		& \citet{cantillo2022fare} & Santiago (Chile) \\
		& \citet{guarda2016behind} & Santiago (Chile) \\
		& \citet{barabino2023evaluating} & Italy\\
		
		\addlinespace
		\multicolumn{3}{l}{\textbf{Rear-door entry} \tiny(associated with higher evasion)} \\
		& \citet{lee2011uncovering} & San Francisco (USA) \\
		
		\addlinespace
		\multicolumn{3}{l}{\textbf{Low-income neighbourhood} \tiny(associated with higher evasion)} \\
		& \citet{cantillo2022fare} & Santiago (Chile) \\
		
		\addlinespace
		\multicolumn{3}{l}{\textbf{Crowding / high occupancy} \tiny(associated with higher evasion)} \\
		& \citet{cantillo2022fare} & Santiago (Chile) \\
		& \citet{guarda2016behind} & Santiago (Chile) \\
		
		\addlinespace
		\multicolumn{3}{l}{\textbf{High boarding volume at stop} \tiny(associated with higher evasion)} \\
		& \citet{guarda2016behind} & Santiago (Chile) \\
		
		\addlinespace
		\multicolumn{3}{l}{\textbf{No off-board payment at stop} \tiny(associated with higher evasion)} \\
		& \citet{cantillo2022fare} & Santiago (Chile) \\
		
		\addlinespace
		\multicolumn{3}{l}{\textbf{Weak metro accessibility} \tiny(associated with higher evasion)} \\
		& \citet{cantillo2022fare} & Santiago (Chile) \\
		
		\midrule
		\multicolumn{3}{l}{\textit{Prices and perceptions}} \\
		\addlinespace
		
		\multicolumn{3}{l}{\textbf{Ticket price} \tiny(higher prices associated with more evasion)} \\
		& \citet{troncoso2017fare} & Santiago (Chile) \\
		& \citet{cools2018identification} & Flanders (Belgium) \\
		
		\addlinespace
		\multicolumn{3}{l}{\textbf{Perceived control probability} \tiny(higher perceived control associated with less evasion)} \\
		& \citet{cools2018identification} & Flanders (Belgium) \\
		
		\bottomrule
	\end{tabular}
	\begin{tablenotes}
		\footnotesize
		\item Notes: Parenthetical remarks summarise the general tendency reported across the cited studies for each attribute. Where multiple studies are listed, this reflects the predominant direction found in the literature (rather than a precise or uniform restatement of each individual study's results).
	\end{tablenotes}
\end{table}

There are several studies from Europe. For instance, \citet{bucciol2013unethical} interviewed passengers using the bus system in Reggio Emilia (Italy). They find that young individuals, males, unemployed persons, and non-European immigrants are more likely to evade payment. Moreover, the results indicate that passengers without tickets tend to take shorter trips and are occasional users of public transport. Applying logistic regression models to 2,200 on-board personal interviews collected among passengers of an Italian public transport company, \citet{barabino2015determinants} identify determinants of potential free-rider behaviour and reach a similar conclusion: fare evasion is more likely among males, young individuals, unemployed persons, students, and those with lower levels of education. Using data from Cagliari (Italy), \citet{barabino2022assessing} show that the intention to evade fares increases among males who travel frequently during the day and among females who are dissatisfied with the service. \citet{barabino2023evaluating}, investigating data of a mid-sized Italian bus company, show that for male, young, less educated, and captive riders the severity of fare evasion increases. Finally, their findings also suggest that more evasion occurs in medium-sized vehicles than in short ones.

Using logistic regression to analyse survey data collected in Flanders, the northern part of Belgium, \citet{cools2018identification} find that age and gender are robust socio-demographic predictors of fare evasion, with young and male travellers exhibiting the highest likelihood of evading fares. Moreover, perceptions of ticket prices and perceived control probability directly influence evasion rates.

Beyond Europe, \citet{lee2011uncovering}, using a survey of over 40,000 customers in San Francisco, show that fare evasion varies by route, time of day, and the availability of rear-door entry. In South America, \citet{troncoso2017fare}, using data from the Santiago (Chile) bus system, show that a 10\% increase in fares raises fare evasion by 2 percentage points. Moreover, \citet{cantillo2022fare}, also analysing the case of Santiago (Chile), where evasion rates differ substantially between metro and bus, find that fare evasion is higher among young men, during evening and night periods, in low-income neighbourhoods, on crowded buses, at bus stops without off-board payment, and in areas with weak accessibility to metro stations. Similarly, \citet{guarda2016behind}, also investigating fare evasion in Santiago, reach comparable conclusions, showing that buses with higher boarding volumes at a given stop and higher occupancy levels are more prone to fare evasion. In line with \citet{barabino2023evaluating}, \citet{guarda2016behind} observe a higher evasion rate in longer vehicles with multiple doors.

\subsection{Inspection Strategies}

There are several aspects to consider when discussing the conduct of fare inspections \citep[see, e.g.,][for a review]{barabino2024fare}. 

One dimension concerns the visibility of enforcement, that is, whether inspections are conducted by uniformed or plainclothes staff. From a theoretical perspective, uniformed inspections increase the perceived probability of detection among passengers, thereby deterring opportunistic fare evasion \citep{becker1968crime}. However, this visibility may also enable strategic adaptation, as passengers can identify and avoid inspectors. Plainclothes inspections, by contrast, limit such behavioural adjustment and may therefore be more effective at detecting actual offenders, albeit at the cost of reduced deterrence. Empirical evidence on the relative detection efficiency of these two strategies remains scarce. \citet{keuchel2020effects} provide one of the few direct comparisons, analysing a natural experiment at Stadtwerke Münster, the inner-city bus operator in Münster, Germany, where ticket inspectors switched from uniforms to civilian clothing in December 2016. Their findings suggest that plainclothes inspections increase detections of passengers travelling without a valid ticket, while reducing cases of forgotten or unvalidated tickets. The authors attribute this pattern to the fact that infrequent travellers may fail to recognise plainclothes inspectors, making evasion detection more likely under covert enforcement. Regarding inspector visibility, \citet{egu2020comparable} caution that inspection records are inherently shaped by enforcement practices, as visible controls may systematically underestimate true evasion rates due to avoidance behaviour by passengers.

Another stream of literature discusses the maximization of inspection efficiency by checking the largest possible number of passengers within a short period of time. However, maximising the number of checked passengers does not necessarily lead to the highest number of detected fare evaders. To address this issue, \citet{barabino2023evaluating} introduce a framework that assigns a risk value to segments of routes as a function of the frequency of fare evasion, its associated severity, and exposure measures. The framework integrates fare evasion determinants, prediction models, and a risk-based assessment method. Using approximately 20,000 real-world inspection records from a mid-sized Italian bus company, the authors demonstrate the applicability of their framework. These considerations motivate our definition of inspection efficiency, defined as the number of detected fare evaders per unit of inspection time (encompassing both in-vehicle and between-vehicle time) rather than the number of passengers checked (see Section~\ref{Data}). Moreover, closely related to our study is the paper of \citet{delfau2018optimization}, who apply reinforcement learning to optimise inspector routing using data from a bus network in the Paris region. Unlike our approach, however, their method does not address the causal effect of inspection strategies.

More broadly related to our study is the literature on determining the optimal number of inspections. This can be addressed empirically. For instance, using data from Italy over a three-year period, \citet{barabino2014fare} develop an economic framework to determine the optimal inspection level, defined as the ratio of inspected passengers to total carried passengers. Their results indicate that the optimal inspection level amounts to 3.8\%. More recently, \citet{barabino2019moving} conclude that the optimal inspection rate, measured over a long time horizon and defined as the rate that maximises profit, lies in the range of 3.4\%--4.0\%. To this end, they apply an economic framework to approximately 57,000 stop-level inspections collected by public transport companies in Cagliari, Italy.

Finally, also more broadly related to our study, is the literature on optimal inspection activity planning. \citet{yin2012trusts} present a system called Tactical Randomization for Urban Security in Transit Systems (TRUST), which computes patrol strategies designed to deter fare evasion while respecting operational constraints. The inspection scheduling problem is modelled as a leader--follower Stackelberg game, that is, a sequential-move game in which inspectors (the leader) commit to a mixed strategy and passengers (the followers) decide whether to evade or comply. More recently, also building on a Stackelberg game framework, \citet{escalona2024fare} propose an inspection strategy based on in-station selective inspections using an unpredictable patrolling schedule, where a specific schedule is selected each day with a given probability.

\section{Data}\label{Data}

Our primary dataset contains 178,686 records of fare inspections conducted on PostAuto buses in Switzerland in 2025. Each observation corresponds to a single inspection event carried out by one inspector, and includes information on the start and end time of the inspection, the public transport line, the stop at which the inspector boarded, the type of inspection, the number of passengers checked, and the number of passengers found travelling without a valid ticket.

From these raw inspection records, we construct observations at the level of line, month, time of day,\footnote{Time of day is categorised into seven slots: morning, morning commute, midday, afternoon, evening commute, evening, and night.} and inspection type. For each observation, we aggregate the total inspection time (encompassing both in-vehicle and between-vehicle time) and the total number of detected fare evaders (where multiple inspectors operated jointly on the same line segment, both inspection time and detected fare evaders are summed across all inspectors). When an inspection spans over multiple times of day, we split both the inspection time and the number of detected fare evaders according to the time share in each time of the day. Inspection efficiency is then defined as the ratio of detected fare evaders to total inspection time.

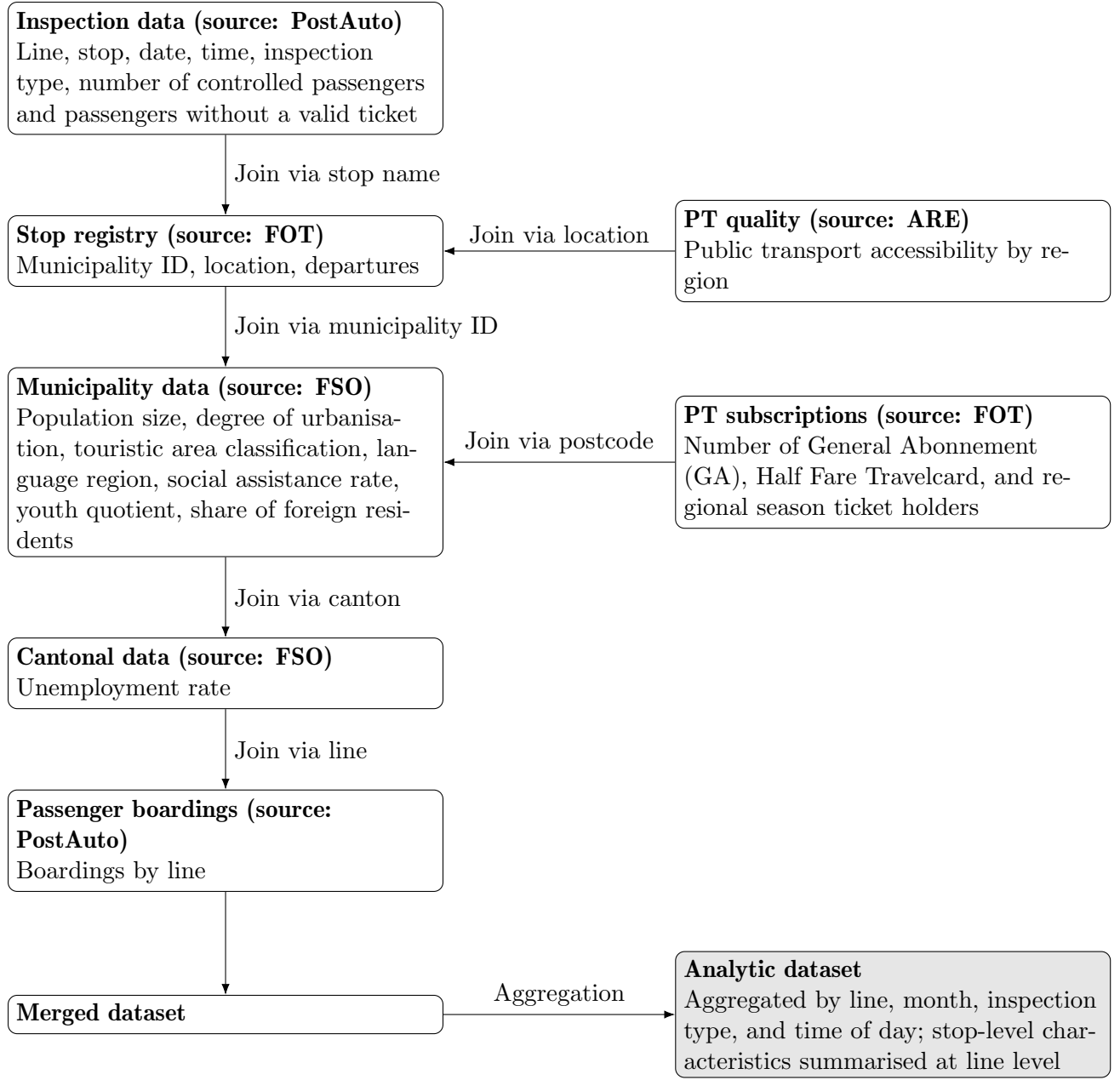
\begin{figure}[ht]
	\centering
	\begin{tikzpicture}[
		node distance=1.25cm and 3.6cm,
		main/.style={draw, rounded corners, align=left, text width=6.5cm},
		highlight/.style={draw, rounded corners, align=left, text width=6.5cm, fill=gray!20},
		arrow/.style={-Latex, line width=0.4pt}
		]
		
		\node[main] (inspection) {\textbf{Inspection data (source: PostAuto)}\\Line, stop, date, time, inspection type, number of controlled passengers and passengers without a valid ticket};
		\node[main, below=of inspection] (stopreg) {\textbf{Stop registry (source: FOT)}\\Municipality ID, location, departures};
		\node[main, right=of stopreg] (Ptquality) {\textbf{PT quality (source: ARE)}\\Public transport accessibility by region};
		\node[main, below=of stopreg] (muni) {\textbf{Municipality data (source: FSO)}\\Population size, degree of urbanisation, touristic area classification, language region, social assistance rate, youth quotient, share of foreign residents};
		\node[main, right=of muni] (PTpass) {\textbf{PT subscriptions (source: FOT)}\\Number of General Abonnement (GA), Half Fare Travelcard, and regional season ticket holders};
		\node[main, below=of muni] (canton) {\textbf{Cantonal data (source: FSO)}\\Unemployment rate};
		\node[main, below=of canton] (boardings) {\textbf{Passenger boardings (source: PostAuto)}\\Boardings by line};
		\node[main, below=1.6cm of boardings] (merged) {\textbf{Merged dataset}};
		\node[highlight, right=of merged] (analytic) {\textbf{Analytic dataset}\\Aggregated by line, month, inspection type, and time of day; stop-level characteristics summarised at line level};
		
		\draw[arrow] (inspection) -- node[right]{Join via stop name} (stopreg);
		\draw[arrow] (Ptquality) -- node[above]{Join via location} (stopreg);
		\draw[arrow] (stopreg) -- node[right]{Join via municipality ID} (muni);
		\draw[arrow] (PTpass) -- node[above]{Join via postcode} (muni);
		\draw[arrow] (muni) -- node[right]{Join via canton} (canton);
		\draw[arrow] (canton) -- node[right]{Join via line} (boardings);
		\draw[arrow] (boardings) -- (merged);
		\draw[arrow] (merged) -- node[above]{Aggregation} (analytic);
		
	\end{tikzpicture}
	\caption{Data enrichment process}
	\label{fig:data_enrichment}
\end{figure}

Based on the stops and lines contained in our inspection dataset, we collected the typical route of each line from the Federal Office of Transport (FOT) via opentransportdata.swiss\footnote{\url{https://data.opentransportdata.swiss/dataset/timetable-2026-gtfs2020}, accessed on May 29, 2026.}, which includes the geolocation of each stop. Using these geolocations together with geodata from swisstopo (swissBOUNDARIES3D\footnote{\url{https://www.swisstopo.admin.ch/de/landschaftsmodell-swissboundaries3d}, accessed on May 29, 2026.}), we mapped each stop to its corresponding municipality (see Figure~\ref{fig:data_enrichment}). This allowed us to merge municipality-level attributes (potentially) associated with fare evasion (see Table~\ref{tab:fare_evasion_attributes}), including population size\footnote{\url{https://mapexplorer.bfs.admin.ch/?obs=main&lang=de\#c=indicator&i=ch_01_02_01a.staendigepop&s=2024&view=map179}, accessed on May 29, 2026.}, degree of urbanisation\footnote{\url{https://www.agvchapp.bfs.admin.ch/de/boundaries?SnapshotDate=01.01.2026&Unit=GDETYP2020}, accessed on May 29, 2026.}, touristic area classification\footnote{\url{https://www.bfs.admin.ch/bfs/de/home/statistiken/kataloge-datenbanken.assetdetail.36217024.html}, accessed on May 29, 2026. Note that this variable does not originate from the literature. However, it may also serve as a proxy for occasional passengers.}, language region\footnote{\url{https://www.bfs.admin.ch/bfs/de/home/statistiken/regionalstatistik/kartengrundlagen/basisgeometrien.assetdetail.33807959.html}, accessed on May 29, 2026.}, social assistance rate, youth quotient, and share of foreign residents, all sourced from the Federal Statistical Office (FSO\footnote{\url{https://mapexplorer.bfs.admin.ch/?obs=main&lang=de\#c=indicator&view=map164}, accessed on May 29, 2026.}). Additionally, we incorporated data from the FOT on the number of General Abonnement (GA), Half Fare Travelcard\footnote{\url{https://data.opentransportdata.swiss/dataset/ga-hta-liste1}, accessed on May 29, 2026.}, and regional season ticket holders\footnote{\url{https://data.opentransportdata.swiss/dataset/verbundsabos}, accessed on May 29, 2026.}, as well as data on public transport accessibility by region from the Federal Office for Spatial Development (ARE\footnote{\url{https://data.geo.admin.ch/browser/index.html\#/collections/ch.are.gueteklassen_oev?.language=de-CH}, accessed on May 29, 2026.}). We further include passenger boardings by line, sourced directly from PostAuto, as a measure of line-level demand. Finally, we include cantonal-level unemployment rates sourced from the FSO\footnote{\url{https://mapexplorer.bfs.admin.ch/?obs=main&lang=de\#c=indicator&view=map164}, accessed on May 29, 2026.}, as municipality-level data are not available for this indicator. 

The (final) analytic dataset is constructed in two steps. First, for each municipality- and stop-level characteristic, we compute summary statistics (namely the mean, median, minimum, maximum, and interquartile range) across all stops belonging to a given line, thereby obtaining line-level contextual characteristics. Second, we aggregate the inspection records by line, month, inspection type, and time of day, summing total inspection time and the number of detected fare evaders to construct the outcome variable. As we also include lagged values of inspection time, we drop the first month of our observations. The aggregated sample comprises  19,882 inspection events (observations) classified as uniformed inspections (\textit{Normalkontrolle}) and 1,466 observations classified as plainclothes inspections (\textit{Präsenzkontrolle}), arising from 824 distinct lines and includes 71 contextual characteristics (see Table \ref{tab:covariates} in Appendix \ref{Appendix_Tabels}).

\FloatBarrier

\section{Identification and estimation}\label{Identification_estimation}

Our empirical strategy proceeds in two steps. First, we apply causal machine learning to estimate the causal effect of plainclothes versus uniformed inspections on inspection efficiency, and to examine whether this effect varies systematically across contextual characteristics. Second, we use optimal policy learning to determine the inspection strategy that maximises inspection efficiency for a given context. Both steps are discussed in the following.

\subsection{Causal machine learning}
We use causal machine learning to estimate the causal effect of inspection strategy on inspection efficiency, and to examine whether this effect varies systematically across contextual characteristics. Specifically, $D$ denotes the binary treatment variable, where $D = 1$ indicates a uniformed inspection (\textit{Präsenzkontrolle}) and $D = 0$ indicates a plainclothes inspection (\textit{Normalkontrolle}). Note that our analysis is conditional on inspections being carried out, meaning that observations without any inspection activity are not included in the sample, and our findings therefore do not generalise to contexts in which no inspection activity takes place. The outcome variable $Y$ captures inspection efficiency, defined as the number of detected fare evaders per unit of inspection time.

Let $Y(1)$ and $Y(0)$ denote the potential outcomes under uniformed and plainclothes inspections, respectively. The average treatment effect (ATE),
\[
\tau = \mathbb{E}[Y(1) - Y(0)],
\]
captures the average effect of uniformed relative to plainclothes inspections across all observations. The conditional average treatment effect (CATE),
\[
\tau(x) = \mathbb{E}[Y(1) - Y(0) \mid X = x],
\]
extends this by capturing how the effect varies across contextual characteristics $X$. The ATE corresponds to the average of the CATE over the distribution of $X$, i.e., $\tau = \mathbb{E}[\tau(X)]$. To examine treatment effect heterogeneity, we further estimate the Sorted Group Average Treatment Effects (GATES),  which quantify the average CATE across $K$ groups ranked by their estimated CATE and test whether these group-specific averages differ significantly from one another.\footnote{Alternatively, GATEs can be estimated for theoretically motivated subgroups defined by specific covariates (e.g., population size, GA ownership, or prior inspection activity), which may yield more directly interpretable results for operational decision-making. For an application of this approach, see \citet{langen2023causal}.} Formally, we can present the GATEs as
\[
\gamma_k = \mathbb{E}[\tau(X) \mid G_k], \quad k = 1, \ldots, K,
\]
where $G_k$ denotes the $k$-th group ranked by the estimated CATE $\hat{\tau}(x)$, also the so-called heterogeneity score. Put simply, observations are sorted from those for whom uniformed inspections are least efficient at detection (G1) to those for whom they are most efficient at detection (G5), based on the estimated CATE. Note that the GATEs do not identify which contextual characteristics drive this heterogeneity, therefore we apply the Best Linear Predictor (BLP) \citep[see][]{semenova2021debiased}, which tests whether the CATE differs systematically across contextual characteristics. Following \citet{semenova2021debiased}, the BLP is implemented by (i) plugging the causal forest predictions into doubly robust scores and (ii) linearly regressing these doubly robust scores on a pre-selected set of contextual characteristics.

The interpretation of our results rests on two assumptions.

\textbf{Assumption 1 (Conditional Independence of Treatment):} Assumption 1 holds if, conditional on observed covariates $X$, the inspection strategy $D$ is independent of potential outcomes. This requires that all factors jointly affecting the choice of inspection strategy and inspection outcomes are observed and included in $X$. Formally,
\[
Y(1), Y(0) \perp D \mid X.
\]

We argue that Assumption 1 is plausible in our setting, based on information provided by PostAuto. Both the decisions of planners and individual inspectors are partly guided by prior inspection results, which are likely correlated with the determinants of fare evasion. The latter is relevant because inspectors often have discretion over their route after completing an inspection. To account for this, we control for prior inspection activity in the same time slot and overall in the preceding month, as well as for the determinants of fare evasion discussed in Section~\ref{Lit_determinants}. Second, coordinated inspections constitute a specific inspection format in which multiple inspectors simultaneously cover all entry and exit points of a vehicle, effectively eliminating passengers' ability to avoid inspection. These inspections are often conducted in uniformed mode and tend to occur at locations where multiple lines can be controlled simultaneously, i.e., where public transport accessibility is higher. Importantly, inspectors participating in coordinated operations often subsequently conduct individual inspections without changing into plainclothes attire, which further increases the share of uniformed inspections in high-accessibility locations. By controlling for public transport accessibility, we further address the possibility that the spatial distribution of coordinated inspections confounds the treatment assignment. (Note that as coordinated inspections differ systematically from standard inspections, we exclude them from the main analysis and revisit them in Section~\ref{Discussion} and Table~\ref{tab:ate_t2}.)

\textbf{Assumption 2 (Common Support):} Assumption 2 requires that, for all $x$ in the support of $X$, both inspection strategies have a strictly positive probability of being assigned — or in other words, substantial overlap in covariates between the treated and control group. Formally,
\[
0 < \mathbb{P}(D = 1 \mid X = x) < 1.
\]
The conditional treatment probability $\mathbb{P}(D = 1 \mid X = x)$ is also referred to as the propensity score. Under Assumptions 1 and 2, the CATE $\tau(x)$ is identified for all $x$ in the support of $X$.

We apply the causal forest (CF) approach by \citet{wager2018estimation} and \citet{athey2019generalized} to estimate the ATE and CATE. By growing many trees, the causal forest extends the random forest algorithm to the estimation of heterogeneous treatment effects by recursively partitioning the covariate space $X$ into subgroups with similar treatment effects, rather than similar outcomes. To avoid overfitting, the algorithm relies on honesty, meaning that separate subsamples are used for determining the tree structure and for estimating treatment effects within leaves. The CATE $\hat{\tau}(x)$ is then obtained by averaging the local treatment effect estimates across all trees. To examine treatment effect heterogeneity across a pre-selected set of contextual characteristics, we estimate the Best Linear Predictor (BLP) of the CATE on these covariates \citep[see, e.g.,][]{Chernozhukov2025}. We further estimate the GATEs to quantify average treatment effects across groups ranked by their estimated CATE. All estimates are obtained using the \texttt{grf} package by \citet{tibshirani2020package} and the \texttt{GenericML} package by \citet{welz2022genericml} for the statistical software \texttt{R}.

\subsection{Optimal policy learning}
Although the causal forest results reveal that plainclothes inspections are, on average, more effective at the detection margin than uniformed inspections, there may nonetheless exist contexts in which uniformed inspections are preferable. Optimal policy learning \citep[see, e.g.,][]{kitagawa2018should} goes beyond effect estimation by directly targeting optimal decision-making: rather than asking what is the effect of a given inspection strategy, it asks which inspection strategy should be assigned to a given context to maximise inspection efficiency. The approach aims to optimally allocate an intervention across subgroups based on the size of their individualized effects, taking into account observed contextual characteristics $X$.

Formally, we seek a policy function $\pi: \mathcal{X} \rightarrow \{0, 1\}$ that maps contextual characteristics $X$ to an inspection strategy, where $\pi(x) = 1$ indicates that uniformed inspections are recommended and $\pi(x) = 0$ indicates that plainclothes inspections are recommended. The optimal policy maximises the expected inspection efficiency,
\[
\pi^* = \arg\max_{\pi} \mathbb{E}[Y(\pi(X))],
\]
where $Y(\pi(X))$ denotes the potential outcome under the strategy assigned by $\pi$. To estimate $\pi^*$, we follow \citet{athey2021policy} and apply a policy tree, which is a shallow decision tree that partitions the covariate space into regions and assigns an inspection strategy to each region. The complexity of the optimal allocation rule is regulated by the maximum number of leaves, which determines how many distinct segments with potentially different inspection strategies are considered. In our application, we set the maximum number of leaves to four, yielding at most four distinct contexts with potentially different inspection strategy recommendations.

The policy tree is fitted using doubly robust scores obtained from the causal forest, which provide an approximately unbiased signal of the individual treatment effect and thereby allow for valid inference on the optimal policy. To ensure interpretability for practitioners, we train the causal forest on the full set of covariates $X$ but restrict the policy tree to a pre-selected set of binary contextual characteristics that are (easier) actionable and easy to communicate. All estimates are obtained using the \texttt{policytree} package by \citet{sverdrup2026policytree} for the statistical software \texttt{R}.

\section{Results}\label{Results}
\subsection{Descriptive analysis}

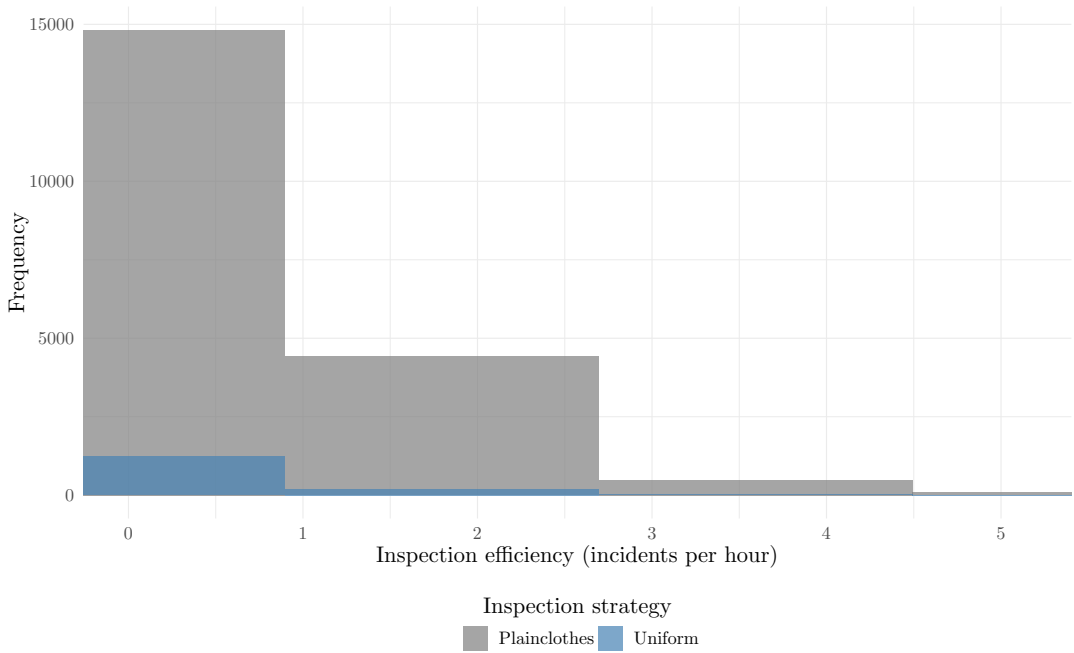
\begin{figure}[H] 
	\centering
	\caption{Distribution of Inspection Efficiency per Line (excluding top 0.5\%)}\label{Inspection_Efficiency}	\resizebox{0.9\textwidth}{!}{
\begin{tikzpicture}[x=1pt,y=1pt]
\definecolor{fillColor}{RGB}{255,255,255}
\path[use as bounding box,fill=fillColor,fill opacity=0.00] (0,0) rectangle (578.16,361.35);
\begin{scope}
\path[clip] (  0.00,  0.00) rectangle (578.16,361.35);
\definecolor{fillColor}{RGB}{255,255,255}

\path[fill=fillColor] (  0.00,  0.00) rectangle (578.16,361.35);
\end{scope}
\begin{scope}
\path[clip] ( 44.91, 82.35) rectangle (572.66,355.85);
\definecolor{drawColor}{gray}{0.92}

\path[draw=drawColor,line width= 0.3pt,line join=round] ( 44.91,136.71) --
	(572.66,136.71);

\path[draw=drawColor,line width= 0.3pt,line join=round] ( 44.91,220.57) --
	(572.66,220.57);

\path[draw=drawColor,line width= 0.3pt,line join=round] ( 44.91,304.43) --
	(572.66,304.43);

\path[draw=drawColor,line width= 0.3pt,line join=round] (115.52, 82.35) --
	(115.52,355.85);

\path[draw=drawColor,line width= 0.3pt,line join=round] (208.75, 82.35) --
	(208.75,355.85);

\path[draw=drawColor,line width= 0.3pt,line join=round] (301.99, 82.35) --
	(301.99,355.85);

\path[draw=drawColor,line width= 0.3pt,line join=round] (395.22, 82.35) --
	(395.22,355.85);

\path[draw=drawColor,line width= 0.3pt,line join=round] (488.46, 82.35) --
	(488.46,355.85);

\path[draw=drawColor,line width= 0.6pt,line join=round] ( 44.91, 94.79) --
	(572.66, 94.79);

\path[draw=drawColor,line width= 0.6pt,line join=round] ( 44.91,178.64) --
	(572.66,178.64);

\path[draw=drawColor,line width= 0.6pt,line join=round] ( 44.91,262.50) --
	(572.66,262.50);

\path[draw=drawColor,line width= 0.6pt,line join=round] ( 44.91,346.35) --
	(572.66,346.35);

\path[draw=drawColor,line width= 0.6pt,line join=round] ( 68.90, 82.35) --
	( 68.90,355.85);

\path[draw=drawColor,line width= 0.6pt,line join=round] (162.13, 82.35) --
	(162.13,355.85);

\path[draw=drawColor,line width= 0.6pt,line join=round] (255.37, 82.35) --
	(255.37,355.85);

\path[draw=drawColor,line width= 0.6pt,line join=round] (348.61, 82.35) --
	(348.61,355.85);

\path[draw=drawColor,line width= 0.6pt,line join=round] (441.84, 82.35) --
	(441.84,355.85);

\path[draw=drawColor,line width= 0.6pt,line join=round] (535.08, 82.35) --
	(535.08,355.85);
\definecolor{fillColor}{RGB}{127,127,127}

\path[fill=fillColor,fill opacity=0.70] (-14.87, 94.79) rectangle (152.66,343.42);

\path[fill=fillColor,fill opacity=0.70] (152.66, 94.79) rectangle (320.19,168.88);

\path[fill=fillColor,fill opacity=0.70] (320.19, 94.79) rectangle (487.72,102.89);

\path[fill=fillColor,fill opacity=0.70] (487.72, 94.79) rectangle (655.25, 96.38);

\path[fill=fillColor,fill opacity=0.70] (655.25, 94.79) rectangle (822.78, 95.42);

\path[fill=fillColor,fill opacity=0.70] (822.78, 94.79) rectangle (990.31, 94.95);

\path[fill=fillColor,fill opacity=0.70] (990.31, 94.79) rectangle (1157.84, 94.90);

\path[fill=fillColor,fill opacity=0.70] (1157.84, 94.79) rectangle (1325.37, 94.82);

\path[fill=fillColor,fill opacity=0.70] (1325.37, 94.79) rectangle (1492.90, 94.80);

\path[fill=fillColor,fill opacity=0.70] (1492.90, 94.79) rectangle (1660.43, 94.80);

\path[fill=fillColor,fill opacity=0.70] (1660.43, 94.79) rectangle (1827.96, 94.79);

\path[fill=fillColor,fill opacity=0.70] (1827.96, 94.79) rectangle (1995.49, 94.79);

\path[fill=fillColor,fill opacity=0.70] (1995.49, 94.79) rectangle (2163.02, 94.79);

\path[fill=fillColor,fill opacity=0.70] (2163.02, 94.79) rectangle (2330.54, 94.79);

\path[fill=fillColor,fill opacity=0.70] (2330.54, 94.79) rectangle (2498.07, 94.80);

\path[fill=fillColor,fill opacity=0.70] (2498.07, 94.79) rectangle (2665.60, 94.79);

\path[fill=fillColor,fill opacity=0.70] (2665.60, 94.79) rectangle (2833.13, 94.79);

\path[fill=fillColor,fill opacity=0.70] (2833.13, 94.79) rectangle (3000.66, 94.79);
\definecolor{fillColor}{RGB}{70,130,180}

\path[fill=fillColor,fill opacity=0.70] (-14.87, 94.79) rectangle (152.66,115.55);

\path[fill=fillColor,fill opacity=0.70] (152.66, 94.79) rectangle (320.19, 98.04);

\path[fill=fillColor,fill opacity=0.70] (320.19, 94.79) rectangle (487.72, 95.21);

\path[fill=fillColor,fill opacity=0.70] (487.72, 94.79) rectangle (655.25, 94.85);

\path[fill=fillColor,fill opacity=0.70] (655.25, 94.79) rectangle (822.78, 94.79);

\path[fill=fillColor,fill opacity=0.70] (822.78, 94.79) rectangle (990.31, 94.87);

\path[fill=fillColor,fill opacity=0.70] (990.31, 94.79) rectangle (1157.84, 94.79);

\path[fill=fillColor,fill opacity=0.70] (1157.84, 94.79) rectangle (1325.37, 94.79);

\path[fill=fillColor,fill opacity=0.70] (1325.37, 94.79) rectangle (1492.90, 94.79);

\path[fill=fillColor,fill opacity=0.70] (1492.90, 94.79) rectangle (1660.43, 94.79);

\path[fill=fillColor,fill opacity=0.70] (1660.43, 94.79) rectangle (1827.96, 94.79);

\path[fill=fillColor,fill opacity=0.70] (1827.96, 94.79) rectangle (1995.49, 94.79);

\path[fill=fillColor,fill opacity=0.70] (1995.49, 94.79) rectangle (2163.02, 94.79);

\path[fill=fillColor,fill opacity=0.70] (2163.02, 94.79) rectangle (2330.54, 94.79);

\path[fill=fillColor,fill opacity=0.70] (2330.54, 94.79) rectangle (2498.07, 94.79);

\path[fill=fillColor,fill opacity=0.70] (2498.07, 94.79) rectangle (2665.60, 94.79);

\path[fill=fillColor,fill opacity=0.70] (2665.60, 94.79) rectangle (2833.13, 94.79);

\path[fill=fillColor,fill opacity=0.70] (2833.13, 94.79) rectangle (3000.66, 94.79);
\end{scope}
\begin{scope}
\path[clip] (  0.00,  0.00) rectangle (578.16,361.35);
\definecolor{drawColor}{gray}{0.30}

\node[text=drawColor,anchor=base east,inner sep=0pt, outer sep=0pt, scale=  0.88] at ( 39.96, 91.76) {0};

\node[text=drawColor,anchor=base east,inner sep=0pt, outer sep=0pt, scale=  0.88] at ( 39.96,175.61) {5000};

\node[text=drawColor,anchor=base east,inner sep=0pt, outer sep=0pt, scale=  0.88] at ( 39.96,259.47) {10000};

\node[text=drawColor,anchor=base east,inner sep=0pt, outer sep=0pt, scale=  0.88] at ( 39.96,343.32) {15000};
\end{scope}
\begin{scope}
\path[clip] (  0.00,  0.00) rectangle (578.16,361.35);
\definecolor{drawColor}{gray}{0.30}

\node[text=drawColor,anchor=base,inner sep=0pt, outer sep=0pt, scale=  0.88] at ( 68.90, 71.34) {0};

\node[text=drawColor,anchor=base,inner sep=0pt, outer sep=0pt, scale=  0.88] at (162.13, 71.34) {1};

\node[text=drawColor,anchor=base,inner sep=0pt, outer sep=0pt, scale=  0.88] at (255.37, 71.34) {2};

\node[text=drawColor,anchor=base,inner sep=0pt, outer sep=0pt, scale=  0.88] at (348.61, 71.34) {3};

\node[text=drawColor,anchor=base,inner sep=0pt, outer sep=0pt, scale=  0.88] at (441.84, 71.34) {4};

\node[text=drawColor,anchor=base,inner sep=0pt, outer sep=0pt, scale=  0.88] at (535.08, 71.34) {5};
\end{scope}
\begin{scope}
\path[clip] (  0.00,  0.00) rectangle (578.16,361.35);
\definecolor{drawColor}{RGB}{0,0,0}

\node[text=drawColor,anchor=base,inner sep=0pt, outer sep=0pt, scale=  1.10] at (308.78, 59.31) {Inspection efficiency (incidents per hour)};
\end{scope}
\begin{scope}
\path[clip] (  0.00,  0.00) rectangle (578.16,361.35);
\definecolor{drawColor}{RGB}{0,0,0}

\node[text=drawColor,rotate= 90.00,anchor=base,inner sep=0pt, outer sep=0pt, scale=  1.10] at ( 13.08,219.10) {Frequency};
\end{scope}
\begin{scope}
\path[clip] (  0.00,  0.00) rectangle (578.16,361.35);
\definecolor{drawColor}{RGB}{0,0,0}

\node[text=drawColor,anchor=base,inner sep=0pt, outer sep=0pt, scale=  1.10] at (308.78, 32.02) {Inspection strategy};
\end{scope}
\begin{scope}
\path[clip] (  0.00,  0.00) rectangle (578.16,361.35);
\definecolor{fillColor}{RGB}{127,127,127}

\path[fill=fillColor,fill opacity=0.70] (247.59, 11.71) rectangle (260.62, 24.74);
\end{scope}
\begin{scope}
\path[clip] (  0.00,  0.00) rectangle (578.16,361.35);
\definecolor{fillColor}{RGB}{70,130,180}

\path[fill=fillColor,fill opacity=0.70] (319.65, 11.71) rectangle (332.68, 24.74);
\end{scope}
\begin{scope}
\path[clip] (  0.00,  0.00) rectangle (578.16,361.35);
\definecolor{drawColor}{RGB}{0,0,0}

\node[text=drawColor,anchor=base west,inner sep=0pt, outer sep=0pt, scale=  0.88] at (266.84, 15.20) {Plainclothes};
\end{scope}
\begin{scope}
\path[clip] (  0.00,  0.00) rectangle (578.16,361.35);
\definecolor{drawColor}{RGB}{0,0,0}

\node[text=drawColor,anchor=base west,inner sep=0pt, outer sep=0pt, scale=  0.88] at (338.89, 15.20) {Uniform};
\end{scope}
\end{tikzpicture}}
\end{figure}

The average inspection efficiency amounts to 0.6 incidents per hour (SD = 1.0). Figure~\ref{Inspection_Efficiency} and Table~\ref{tab:desc_stats_by_strategy} present the distribution of inspection efficiency and selected covariate means by inspection strategy for the analytic sample. As shown in Figure~\ref{Inspection_Efficiency}, the distribution is right-skewed for both inspection strategies, with the majority of inspections yielding zero or near-zero detections.

Plainclothes inspections yield a higher average inspection efficiency (0.61 incidents per hour) compared to uniformed inspections (0.41 incidents per hour), providing a first descriptive indication that plainclothes inspections may be more effective at detecting fare evaders. This raw difference, however, does not account for confounding factors, which motivates the causal machine learning approach presented in Section~\ref{Identification_estimation}.

On average, lines assigned uniformed inspections tend to have larger populations (16,253 vs.\ 12,452 inhabitants), higher GA ownership rates (0.07 vs.\ 0.05), and higher half-fare card ownership rates (0.50 vs.\ 0.38), suggesting that uniformed inspections are more frequently conducted on lines serving urban areas with lower unemployment rates (0.02 vs. \ 0.03). This is in line with the higher share of very good PT accessibility (0.13 vs, \ 0.10). The share of foreign residents is somewhat lower for uniformed inspections (0.18 vs.\ 0.22), while social assistance rates, and youth dependency ratios are broadly similar across both strategies. Regarding prior inspection activity, lines assigned uniformed inspections show somewhat higher total inspection hours (both plainclothes and uniform) in the prior month (17.1 vs.\ 13.6 hours) as well as slightly higher inspection hours in the same time slot (2.8 vs.\ 2.3 hours), suggesting that uniformed inspections tend to be deployed on lines with a higher baseline level of inspection activity. Standardised differences indicate meaningful imbalance (${|d| > 0.1}$) for all variables (see Table~\ref{tab:desc_stats_by_strategy}).

\subsection{Causal machine learning}

\subsubsection{Average treatment effect}
Table~\ref{tab:ate} reports the estimated average treatment effect (ATE) of uniformed relative to plainclothes inspections on inspection efficiency. Given the substantial imbalance between inspection strategies in the analytic sample (6.9\% uniformed vs.\ 93.1\% plainclothes), we use the overlap target sample as our main specification, which upweights observations with propensity scores close to 0.5 and thereby reduces sensitivity to regions of limited common support (see Figure~\ref{fig:pscore} in the Appendix).\footnote{The distribution of estimated propensity scores reported in Figures~\ref{fig:pscore} and~\ref{fig:pscore_uniformed} reveals that the vast majority of uniformed inspections are concentrated near zero, reflecting the strong imbalance in the analytic sample. Nevertheless, a non-trivial share of uniformed observations exhibit propensity scores in the intermediate range, (potentially) providing sufficient overlap for identification. This should be kept in mind when interpreting the results under Assumption~2 (Common Support).} The causal forest yields an ATE of $-$0.156 incidents per hour (SE = 0.027, $p < 0.001$), indicating that uniformed inspections detect, on average, 0.156 fewer fare evaders per inspection hour compared to plainclothes inspections. This effect is statistically significant at the 1\% level and economically meaningful: given that the sample average inspection efficiency amounts to 0.6 incidents per hour, it implies a relative reduction of approximately 26\%. In summary, the results suggest that, on average, plainclothes inspections are more effective at detecting fare evaders than uniformed inspections. In the following subsection, we examine whether this average effect masks systematic heterogeneity across contextual characteristics.

\begin{table}[H]
	\centering
	\caption{Descriptive statistics by inspection strategy}
	\label{tab:desc_stats_by_strategy}
	\begin{tabular}{lccccc}
		\hline
		& \multicolumn{2}{c}{Plainclothes} & \multicolumn{2}{c}{Uniformed} & \\
		\cmidrule(lr){2-3} \cmidrule(lr){4-5}
		Variable & Mean & SD & Mean & SD & Std. diff. \\
		\hline
		\textit{Outcome} & & & & & \\
		Inspection efficiency (incidents per hour) & 0.613 & 1.010 & 0.413 & 0.861 & 0.214 \\
		\hline
		\textit{Sociodemographic characteristics} & & & & & \\
		Population & 12,452 & 19,698 & 16,253 & 26,143 & $-$0.164 \\
		Share of foreign population & 0.222 & 0.071 & 0.175 & 0.065 & 0.699 \\
		Youth dependency ratio & 0.333 & 0.037 & 0.319 & 0.031 & 0.408 \\
		Social assistance rate & 0.021 & 0.013 & 0.023 & 0.011 & $-$0.180 \\
		Unemployment rate & 0.027 & 0.009 & 0.021 & 0.004 & 0.750 \\
		\hline
		\textit{Public transport characteristics} & & & & & \\
		GA ownership rate & 0.048 & 0.059 & 0.072 & 0.109 & $-$0.280 \\
		Half-fare card ownership rate & 0.384 & 0.336 & 0.496 & 0.649 & $-$0.216 \\
		PT accessibility (class A) & 0.097 & 0.110 & 0.132 & 0.158 & $-$0.255 \\
		\hline
		\textit{Prior inspection activity} & & & & & \\
		Inspection hours (same time slot, prior month) & 2.3 & 3.5 & 2.8 & 4.3 & $-$0.136 \\
		Inspection hours (total, prior month) & 13.6 & 15.9 & 17.1 & 19.3 & $-$0.196 \\
		\hline
		Observations & \multicolumn{2}{c}{19,882} & \multicolumn{2}{c}{1,466} & \\
		\hline
	\end{tabular}
	\begin{tablenotes}
		\small
		\item Notes: Inspection efficiency is measured as incidents per hour. Covariates are line-level averages constructed from stop-level information. Prior inspection activity refers to the total inspection hours recorded in the previous month, either across all time slots or restricted to the same time slot as the current observation. Day time variables (commute evening, evening) are omitted as both groups show zero variation. SD denotes standard deviation. We use the standardized difference to assess imbalance \citep{rosenbaum1985constructing}, interpreting values of $|d| > 0.1$ as indicative of imbalance.
	\end{tablenotes}
\end{table}

\begin{table}[H]
	\centering
	\caption{Average Treatment Effect of Uniformed versus Plainclothes Inspections}
	\label{tab:ate}
	\begin{tabular}{lccc}
		\hline
		& Estimate & Std. Error & $p$-value \\
		\hline
		Causal Forest (overlap) & $-$0.156 & 0.027 & $<$0.001 \\
		\hline
		Number of observations & \multicolumn{3}{c}{21,100} \\
		\hline
	\end{tabular}
	\begin{minipage}{0.7\textwidth}
		\footnotesize\textit{Notes: The average treatment effect is estimated using the causal forest with target sample set to the overlap population ($D = 1$ = uniformed, $D = 0$ = plainclothes). The $p$-value is based on a two-sided $z$-test. Results using the treated target sample are reported in Table~\ref{tab:ate_robustness} in the Appendix. Only complete cases are used for the estimation.}
	\end{minipage}
\end{table}

\subsubsection{Treatment effect heterogeneity}

\begin{figure}[H]
	\centering
	\caption{Distribution of Estimated CATEs}\label{Dist_effect}
	\resizebox{0.9\textwidth}{!}{
\begin{tikzpicture}[x=1pt,y=1pt]
\definecolor{fillColor}{RGB}{255,255,255}
\path[use as bounding box,fill=fillColor,fill opacity=0.00] (0,0) rectangle (578.16,361.35);
\begin{scope}
\path[clip] (  0.00,  0.00) rectangle (578.16,361.35);
\definecolor{fillColor}{RGB}{255,255,255}

\path[fill=fillColor] (  0.00,  0.00) rectangle (578.16,361.35);
\end{scope}
\begin{scope}
\path[clip] ( 40.51, 30.69) rectangle (572.66,355.85);
\definecolor{drawColor}{gray}{0.92}

\path[draw=drawColor,line width= 0.3pt,line join=round] ( 40.51, 82.28) --
	(572.66, 82.28);

\path[draw=drawColor,line width= 0.3pt,line join=round] ( 40.51,155.90) --
	(572.66,155.90);

\path[draw=drawColor,line width= 0.3pt,line join=round] ( 40.51,229.53) --
	(572.66,229.53);

\path[draw=drawColor,line width= 0.3pt,line join=round] ( 40.51,303.15) --
	(572.66,303.15);

\path[draw=drawColor,line width= 0.3pt,line join=round] ( 47.42, 30.69) --
	( 47.42,355.85);

\path[draw=drawColor,line width= 0.3pt,line join=round] ( 81.98, 30.69) --
	( 81.98,355.85);

\path[draw=drawColor,line width= 0.3pt,line join=round] (116.53, 30.69) --
	(116.53,355.85);

\path[draw=drawColor,line width= 0.3pt,line join=round] (151.09, 30.69) --
	(151.09,355.85);

\path[draw=drawColor,line width= 0.3pt,line join=round] (185.64, 30.69) --
	(185.64,355.85);

\path[draw=drawColor,line width= 0.3pt,line join=round] (220.20, 30.69) --
	(220.20,355.85);

\path[draw=drawColor,line width= 0.3pt,line join=round] (254.75, 30.69) --
	(254.75,355.85);

\path[draw=drawColor,line width= 0.3pt,line join=round] (289.31, 30.69) --
	(289.31,355.85);

\path[draw=drawColor,line width= 0.3pt,line join=round] (323.86, 30.69) --
	(323.86,355.85);

\path[draw=drawColor,line width= 0.3pt,line join=round] (358.42, 30.69) --
	(358.42,355.85);

\path[draw=drawColor,line width= 0.3pt,line join=round] (392.97, 30.69) --
	(392.97,355.85);

\path[draw=drawColor,line width= 0.3pt,line join=round] (427.53, 30.69) --
	(427.53,355.85);

\path[draw=drawColor,line width= 0.3pt,line join=round] (462.08, 30.69) --
	(462.08,355.85);

\path[draw=drawColor,line width= 0.3pt,line join=round] (496.64, 30.69) --
	(496.64,355.85);

\path[draw=drawColor,line width= 0.3pt,line join=round] (531.19, 30.69) --
	(531.19,355.85);

\path[draw=drawColor,line width= 0.3pt,line join=round] (565.75, 30.69) --
	(565.75,355.85);

\path[draw=drawColor,line width= 0.6pt,line join=round] ( 40.51, 45.47) --
	(572.66, 45.47);

\path[draw=drawColor,line width= 0.6pt,line join=round] ( 40.51,119.09) --
	(572.66,119.09);

\path[draw=drawColor,line width= 0.6pt,line join=round] ( 40.51,192.72) --
	(572.66,192.72);

\path[draw=drawColor,line width= 0.6pt,line join=round] ( 40.51,266.34) --
	(572.66,266.34);

\path[draw=drawColor,line width= 0.6pt,line join=round] ( 40.51,339.97) --
	(572.66,339.97);

\path[draw=drawColor,line width= 0.6pt,line join=round] ( 64.70, 30.69) --
	( 64.70,355.85);

\path[draw=drawColor,line width= 0.6pt,line join=round] ( 99.25, 30.69) --
	( 99.25,355.85);

\path[draw=drawColor,line width= 0.6pt,line join=round] (133.81, 30.69) --
	(133.81,355.85);

\path[draw=drawColor,line width= 0.6pt,line join=round] (168.36, 30.69) --
	(168.36,355.85);

\path[draw=drawColor,line width= 0.6pt,line join=round] (202.92, 30.69) --
	(202.92,355.85);

\path[draw=drawColor,line width= 0.6pt,line join=round] (237.47, 30.69) --
	(237.47,355.85);

\path[draw=drawColor,line width= 0.6pt,line join=round] (272.03, 30.69) --
	(272.03,355.85);

\path[draw=drawColor,line width= 0.6pt,line join=round] (306.59, 30.69) --
	(306.59,355.85);

\path[draw=drawColor,line width= 0.6pt,line join=round] (341.14, 30.69) --
	(341.14,355.85);

\path[draw=drawColor,line width= 0.6pt,line join=round] (375.70, 30.69) --
	(375.70,355.85);

\path[draw=drawColor,line width= 0.6pt,line join=round] (410.25, 30.69) --
	(410.25,355.85);

\path[draw=drawColor,line width= 0.6pt,line join=round] (444.81, 30.69) --
	(444.81,355.85);

\path[draw=drawColor,line width= 0.6pt,line join=round] (479.36, 30.69) --
	(479.36,355.85);

\path[draw=drawColor,line width= 0.6pt,line join=round] (513.92, 30.69) --
	(513.92,355.85);

\path[draw=drawColor,line width= 0.6pt,line join=round] (548.47, 30.69) --
	(548.47,355.85);
\definecolor{fillColor}{RGB}{89,89,89}

\path[fill=fillColor,fill opacity=0.70] ( 81.98, 45.47) rectangle ( 99.25, 45.91);

\path[fill=fillColor,fill opacity=0.70] ( 99.25, 45.47) rectangle (116.53, 46.35);

\path[fill=fillColor,fill opacity=0.70] (116.53, 45.47) rectangle (133.81, 52.31);

\path[fill=fillColor,fill opacity=0.70] (133.81, 45.47) rectangle (151.09, 68.36);

\path[fill=fillColor,fill opacity=0.70] (151.09, 45.47) rectangle (168.36, 76.83);

\path[fill=fillColor,fill opacity=0.70] (168.36, 45.47) rectangle (185.64, 90.30);

\path[fill=fillColor,fill opacity=0.70] (185.64, 45.47) rectangle (202.92,129.55);

\path[fill=fillColor,fill opacity=0.70] (202.92, 45.47) rectangle (220.20,194.63);

\path[fill=fillColor,fill opacity=0.70] (220.20, 45.47) rectangle (237.47,288.06);

\path[fill=fillColor,fill opacity=0.70] (237.47, 45.47) rectangle (254.75,341.07);

\path[fill=fillColor,fill opacity=0.70] (254.75, 45.47) rectangle (272.03,336.43);

\path[fill=fillColor,fill opacity=0.70] (272.03, 45.47) rectangle (289.31,246.68);

\path[fill=fillColor,fill opacity=0.70] (289.31, 45.47) rectangle (306.59,145.89);

\path[fill=fillColor,fill opacity=0.70] (306.59, 45.47) rectangle (323.86, 84.27);

\path[fill=fillColor,fill opacity=0.70] (323.86, 45.47) rectangle (341.14, 67.04);

\path[fill=fillColor,fill opacity=0.70] (341.14, 45.47) rectangle (358.42, 53.93);

\path[fill=fillColor,fill opacity=0.70] (358.42, 45.47) rectangle (375.70, 50.25);

\path[fill=fillColor,fill opacity=0.70] (375.70, 45.47) rectangle (392.97, 48.56);

\path[fill=fillColor,fill opacity=0.70] (392.97, 45.47) rectangle (410.25, 47.67);

\path[fill=fillColor,fill opacity=0.70] (410.25, 45.47) rectangle (427.53, 46.57);

\path[fill=fillColor,fill opacity=0.70] (427.53, 45.47) rectangle (444.81, 45.76);

\path[fill=fillColor,fill opacity=0.70] (444.81, 45.47) rectangle (462.08, 46.50);

\path[fill=fillColor,fill opacity=0.70] (462.08, 45.47) rectangle (479.36, 45.69);

\path[fill=fillColor,fill opacity=0.70] (479.36, 45.47) rectangle (496.64, 45.83);

\path[fill=fillColor,fill opacity=0.70] (496.64, 45.47) rectangle (513.92, 45.54);

\path[fill=fillColor,fill opacity=0.70] (513.92, 45.47) rectangle (531.19, 45.54);

\path[fill=fillColor,fill opacity=0.70] (531.19, 45.47) rectangle (548.47, 45.54);
\end{scope}
\begin{scope}
\path[clip] (  0.00,  0.00) rectangle (578.16,361.35);
\definecolor{drawColor}{gray}{0.30}

\node[text=drawColor,anchor=base east,inner sep=0pt, outer sep=0pt, scale=  0.88] at ( 35.56, 42.44) {0};

\node[text=drawColor,anchor=base east,inner sep=0pt, outer sep=0pt, scale=  0.88] at ( 35.56,116.06) {1000};

\node[text=drawColor,anchor=base east,inner sep=0pt, outer sep=0pt, scale=  0.88] at ( 35.56,189.69) {2000};

\node[text=drawColor,anchor=base east,inner sep=0pt, outer sep=0pt, scale=  0.88] at ( 35.56,263.31) {3000};

\node[text=drawColor,anchor=base east,inner sep=0pt, outer sep=0pt, scale=  0.88] at ( 35.56,336.94) {4000};
\end{scope}
\begin{scope}
\path[clip] (  0.00,  0.00) rectangle (578.16,361.35);
\definecolor{drawColor}{gray}{0.30}

\node[text=drawColor,anchor=base,inner sep=0pt, outer sep=0pt, scale=  0.88] at ( 64.70, 19.68) {-0.70};

\node[text=drawColor,anchor=base,inner sep=0pt, outer sep=0pt, scale=  0.88] at ( 99.25, 19.68) {-0.60};

\node[text=drawColor,anchor=base,inner sep=0pt, outer sep=0pt, scale=  0.88] at (133.81, 19.68) {-0.50};

\node[text=drawColor,anchor=base,inner sep=0pt, outer sep=0pt, scale=  0.88] at (168.36, 19.68) {-0.40};

\node[text=drawColor,anchor=base,inner sep=0pt, outer sep=0pt, scale=  0.88] at (202.92, 19.68) {-0.30};

\node[text=drawColor,anchor=base,inner sep=0pt, outer sep=0pt, scale=  0.88] at (237.47, 19.68) {-0.20};

\node[text=drawColor,anchor=base,inner sep=0pt, outer sep=0pt, scale=  0.88] at (272.03, 19.68) {-0.10};

\node[text=drawColor,anchor=base,inner sep=0pt, outer sep=0pt, scale=  0.88] at (306.59, 19.68) {0.00};

\node[text=drawColor,anchor=base,inner sep=0pt, outer sep=0pt, scale=  0.88] at (341.14, 19.68) {0.10};

\node[text=drawColor,anchor=base,inner sep=0pt, outer sep=0pt, scale=  0.88] at (375.70, 19.68) {0.20};

\node[text=drawColor,anchor=base,inner sep=0pt, outer sep=0pt, scale=  0.88] at (410.25, 19.68) {0.30};

\node[text=drawColor,anchor=base,inner sep=0pt, outer sep=0pt, scale=  0.88] at (444.81, 19.68) {0.40};

\node[text=drawColor,anchor=base,inner sep=0pt, outer sep=0pt, scale=  0.88] at (479.36, 19.68) {0.50};

\node[text=drawColor,anchor=base,inner sep=0pt, outer sep=0pt, scale=  0.88] at (513.92, 19.68) {0.60};

\node[text=drawColor,anchor=base,inner sep=0pt, outer sep=0pt, scale=  0.88] at (548.47, 19.68) {0.70};
\end{scope}
\begin{scope}
\path[clip] (  0.00,  0.00) rectangle (578.16,361.35);
\definecolor{drawColor}{RGB}{0,0,0}

\node[text=drawColor,anchor=base,inner sep=0pt, outer sep=0pt, scale=  1.10] at (306.59,  7.64) {Estimated Individual Treatment Effects (incidents per hour)};
\end{scope}
\begin{scope}
\path[clip] (  0.00,  0.00) rectangle (578.16,361.35);
\definecolor{drawColor}{RGB}{0,0,0}

\node[text=drawColor,rotate= 90.00,anchor=base,inner sep=0pt, outer sep=0pt, scale=  1.10] at ( 13.08,193.27) {Frequency};
\end{scope}
\end{tikzpicture}}
\end{figure}
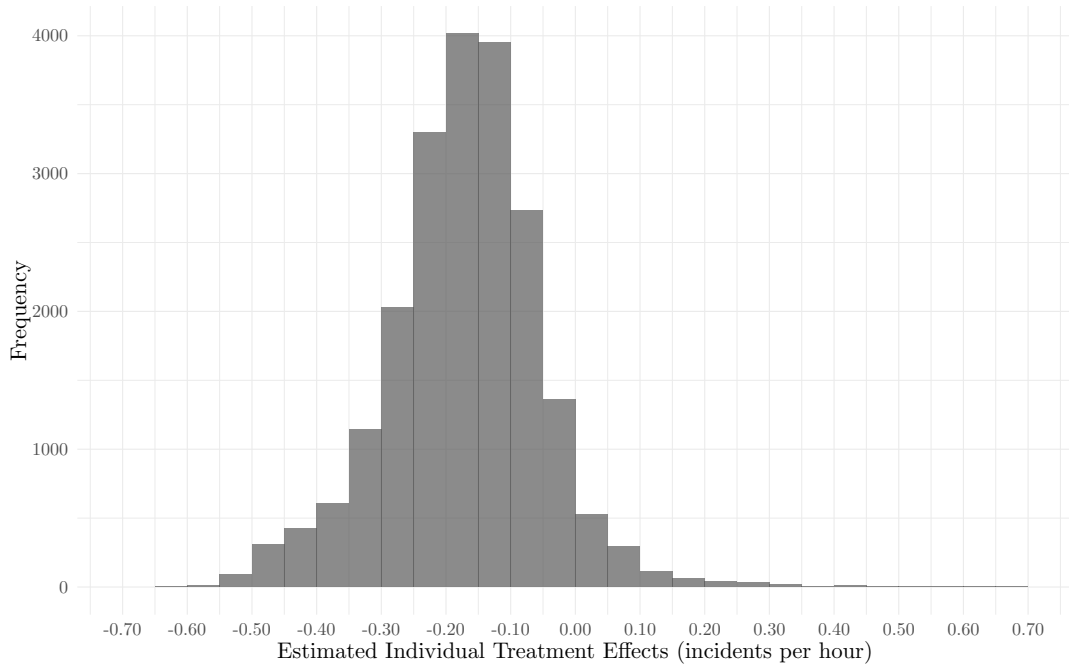

Figure~\ref{Dist_effect} presents the distribution of the estimated CATEs $\hat{\tau}(x)$ across all observations in the analytic sample. The distribution is centred around the negative ATE of $-$0.156 incidents per hour reported in Table~\ref{tab:ate}, with the majority of estimated CATEs being negative. A non-negligible share of observations displays positive estimated CATEs, suggesting that uniformed inspections may be more efficient at detection than plainclothes inspections in certain contexts. However, estimated CATE distributions may appear dispersed even under homogeneous true effects due to estimation uncertainty \citep[see, e.g.,][]{knaus2021machine}, and the observed variation should therefore not be interpreted as evidence of substantial treatment effect heterogeneity per se.

Next, we assess which contextual characteristics are most strongly associated with treatment effect heterogeneity. Table~\ref{tab:var_imp} reports the ten most important predictors of the estimated CATE $\hat{\tau}(x)$, based on the variable importance measure of the causal forest. This measure captures how frequently a variable is used for splitting across all trees in the forest, weighted by the depth at which the split occurs, such that splits higher up in the tree receive greater weight. The resulting scores are normalised to sum to one \citep{athey2019generalized}.

Population size (75th percentile) emerges as the most important predictor, followed by the share of foreign residents (maximum) and GA ownership rates. Half-fare card ownership, total prior inspection hours and the share of very good public transport accessibility also appear among the top predictors. Overall, the results suggest that line-level demand potential (proxied by population size and public transport subscription rates) are the primary drivers of heterogeneity in the effect of uniformed relative to plainclothes inspections on inspection efficiency, while prior inspection activity plays a comparatively minor role (given the other information available in the data).

\begin{table}[H]
	\centering
	\caption{Variable Importance for Predicting Treatment Effect Heterogeneity}
	\label{tab:var_imp}
	\begin{tabular}{lc}
		\hline
		Variable & Importance \\
		\hline
		Population (p75)                              & 0.076 \\
		Share of foreign population (max)             & 0.063 \\
		GA ownership (p25)                            & 0.061 \\
		GA ownership (median)                         & 0.051 \\
		GA ownership (min)                            & 0.037 \\
		GA ownership (p75)                            & 0.036 \\
		Half-fare card ownership (p75)                & 0.036 \\
		Share of foreign population (p75)             & 0.033 \\
		Inspection hours, total (prior month)         & 0.033 \\
		PT accessibility (class A)                    & 0.029 \\
		\hline
	\end{tabular}
	\begin{minipage}{0.6\textwidth}
		\footnotesize\textit{Notes: Variable importance is based on the causal forest and measures the frequency with which a variable is used for splitting, weighted by split depth. Only the ten most important variables are reported.}
	\end{minipage}
\end{table}

We further investigate treatment effect heterogeneity across a pre-selected set of contextual characteristics that are theoretically motivated by the fare evasion literature. Specifically, we include population size as a proxy for line-level demand potential, GA ownership rate as an indicator of the socioeconomic composition of passengers and their likelihood of holding a valid ticket and associated with non-occasional public transport use, the share of foreign residents as a sociodemographic characteristic,\footnote{\label{fn:foreign}The share of foreign residents likely captures more than just migration status per se, but also proxies for socioeconomic status and other sociodemographic characteristics that tend to correlate with nationality at the line level. Moreover, to the extent that inspection strategies differ in their visibility, the share of foreign residents may also reflect potential inspector bias: \citet{mujcic2021colour} provide field experimental evidence from public buses in Australia that minority customers receive systematically less favourable treatment from bus drivers, suggesting that the race or perceived origin of passengers may influence inspector behaviour independently of actual fare evasion propensity.} and total prior inspection hours as well as inspection hours in the same time slot in the prior month as measures of baseline inspection activity that may affect the perceived probability of being controlled (see also Section~\ref{Litrev}).

Table~\ref{tab:blp} reports the Best Linear Predictor (BLP) estimates. Two of the pre-selected covariates are statistically significant at conventional levels, suggesting that the effect of uniformed relative to plainclothes inspections on inspection efficiency varies systematically across GA ownership and the share of foreign population. Overall, the BLP results are consistent with the negative and largely homogeneous ATE reported in Table~\ref{tab:ate}, reinforcing the conclusion that plainclothes inspections tend to exhibit higher detection efficiency than uniformed inspections across a broad range of contextual conditions.

\begin{table}[H]
	\centering
	\caption{Best Linear Predictor (BLP) of the Conditional Average Treatment Effect}
	\label{tab:blp}
	\begin{tabular}{lccc}
		\hline
		& Estimate & Std. Error & $p$-value \\
		\hline
		(Intercept)                               & 0.059 & 0.075 & 0.426 \\
		Population (mean, in 10,000s)             & $-$0.021 & 0.013 & 0.109 \\
		GA ownership (mean)                       & $-$0.323 & 0.158 & 0.041 \\
		Share of foreign population (mean)        & $-$0.009 & 0.004 & 0.025 \\
		Inspection hours, total (prior month)     & $-$0.001 & 0.003 & 0.778 \\
		Inspection hours, same slot (prior month) & 0.002 & 0.010 & 0.825 \\
		\hline
	\end{tabular}
	\begin{minipage}{0.75\textwidth}
		\footnotesize\textit{Notes: Best linear projection of the CATE on pre-selected covariates, estimated using the causal forest with target sample set to the overlap population. Standard errors are cluster- and heteroskedasticity-robust (HC3).}
	\end{minipage}
\end{table}

Table~\ref{tab:gates} reports the Sorted Group Average Treatment Effects (GATES), where observations are sorted into five groups based on their estimated CATE $\hat{\tau}(x)$, ranging from the group with the most negative estimated effect (G1) to the group with the least negative estimated effect (G5). Across all five groups, the GATEs are negative, reinforcing the conclusion that uniformed inspections are less efficient than plainclothes inspections in most contexts. The most affected group (G1) exhibits a GATE of $-$0.34 incidents per hour, while the least affected group (G5) shows a GATE of $-$0.03 incidents per hour, which is not statistically significant different from zero. The difference between G5 and G1 amounts to 0.31 incidents per hour, and a formal test confirms that this difference is statistically significant ($p < 0.001$), suggesting some variation in the magnitude of the effect across contexts, though the direction remains consistently negative throughout the five groups.

\begin{table}[H]
	\centering
	\caption{Sorted Group Average Treatment Effects (GATES)}
	\label{tab:gates}
	\begin{tabular}{lccc}
		\hline
		& Estimate & Std. Error & 95\% CI \\
		\hline
		G1 (lowest $\hat{\tau}$)  & $-$0.338 & 0.066 & [$-$0.468, $-$0.208] \\
		G2                        & $-$0.246 & 0.070 & [$-$0.383, $-$0.109] \\
		G3                        & $-$0.092 & 0.075 & [$-$0.240, $~~$0.056] \\
		G4                        & $-$0.101 & 0.049 & [$-$0.198, $-$0.004] \\
		G5 (highest $\hat{\tau}$) & $-$0.028 & 0.045 & [$-$0.116, $~~$0.061] \\
		\hline
		$\gamma_5 - \gamma_1$ & $~~$0.310 & 0.080 & \\
		\hline
	\end{tabular}
	\begin{minipage}{0.65\textwidth}
		\footnotesize\textit{Notes: Observations are sorted into five groups based on the estimated CATE $\hat{\tau}(x)$ from the causal forest. G1 contains the 20\% of observations with the lowest estimated treatment effect, G5 the 20\% with the highest. Std. Error denotes standard error. The difference between G5 and G1 is statistically significant ($z = 3.88$, $p < 0.001$).}
	\end{minipage}
\end{table}

\subsection{Optimal Policy Learning}
The policy tree is estimated in two stages. In the first stage, a causal forest is trained on the full set of covariates to obtain doubly robust scores, which serve as the welfare-relevant outcome for the subsequent policy optimisation. In the second stage, the policy tree is fitted using only a pre-selected set of interpretable binary covariates: i.e., the medians of population size, GA ownership rate and the share of foreign residents as well as total prior inspection hours, and prior inspection hours in the same time slot.

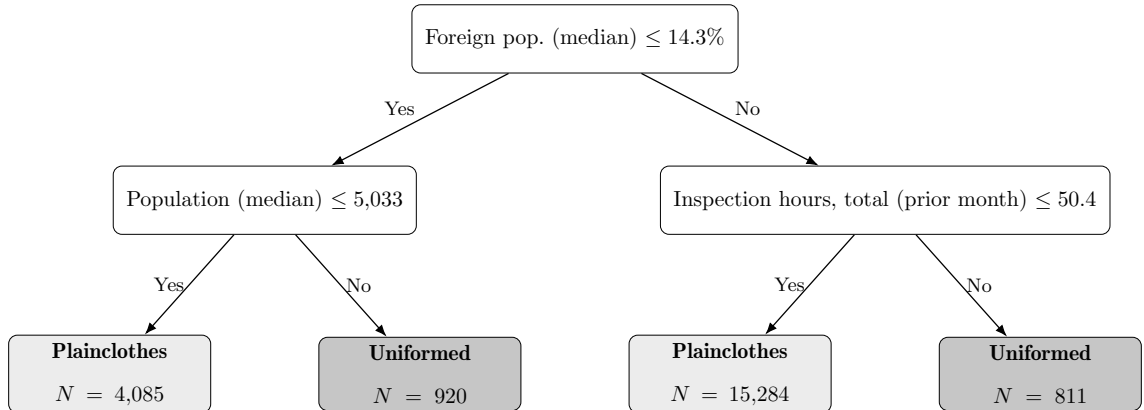
\begin{figure}[H]
	\centering
	\caption{Optimal Inspection Policy Tree}\label{fig:policy_tree}
	\vspace{0.3cm}
	\resizebox{0.95\textwidth}{!}{
		\begin{tikzpicture}[
			decision/.style = {draw, rectangle, rounded corners, align = center,
				inner sep = 6pt, minimum height = 1.1cm, font = \small},
			leaf_plain/.style = {draw, rectangle, rounded corners, align = center,
				text width = 3cm, minimum height = 1.1cm, fill = gray!15, font = \small},
			leaf_uni/.style = {draw, rectangle, rounded corners, align = center,
				text width = 3cm, minimum height = 1.1cm, fill = gray!45, font = \small},
			arrow/.style = {-Latex, line width = 0.6pt},
			lbl/.style = {font = \footnotesize, inner sep = 2pt}
			]
			
			\node[decision] (n1) at (0,0) {Foreign pop.\ (median) $\leq 14.3\%$};
			
			\node[decision] (n2) at (-5,-2.6) {Population (median) $\leq 5{,}033$};
			\node[decision] (n3) at ( 5,-2.6) {Inspection hours, total (prior month) $\leq 50.4$};
			
			\node[leaf_plain] (l1) at (-7.5,-5.4) {\textbf{Plainclothes}\\$N = 4{,}085$};
			\node[leaf_uni]   (l2) at (-2.5,-5.4) {\textbf{Uniformed}\\$N = 920$};
			\node[leaf_plain] (l3) at ( 2.5,-5.4) {\textbf{Plainclothes}\\$N = 15{,}284$};
			\node[leaf_uni]   (l4) at ( 7.5,-5.4) {\textbf{Uniformed}\\$N = 811$};
			
			\draw[arrow] (n1) -- node[lbl, above left]  {Yes} (n2);
			\draw[arrow] (n1) -- node[lbl, above right] {No}  (n3);
			\draw[arrow] (n2) -- node[lbl, left]        {Yes} (l1);
			\draw[arrow] (n2) -- node[lbl, right]       {No}  (l2);
			\draw[arrow] (n3) -- node[lbl, left]        {Yes} (l3);
			\draw[arrow] (n3) -- node[lbl, right]       {No}  (l4);
			
		\end{tikzpicture}
	}
	\vspace{0.4cm}
	\begin{minipage}{0.9\textwidth}
		\footnotesize\setstretch{1.0}\textit{Notes: The policy tree is estimated using doubly robust scores from the causal forest trained on the full set of covariates. The tree is restricted to splits based on median values of pre-selected contextual characteristics. Plainclothes = Normal ($D = 0$), Uniformed = Pr\"{a}senz ($D = 1$). $N$ denotes the number of observations assigned to each leaf. Inspection hours refer to the total across all inspection types on the line in the preceding month.}
	\end{minipage}
\end{figure}

Figure~\ref{fig:policy_tree} presents the optimal inspection strategy as suggested by the policy tree outlined in Section~\ref{Identification_estimation}. The first split is based on the share of foreign residents: for lines serving municipalities with a median share of foreign residents below or equal to 14.3\%, the tree continues to split on the median population size along the line. Lines with a median population of 5,033 or less should be assigned plainclothes inspections ($N = 4{,}085$), while lines with a larger population should be assigned uniformed inspections ($N = 920$). For lines with a median share of foreign residents above 14.3\%, the tree continues to split by total inspection hours in the past month. Lines with 50.4 or less inspection hours in the past month should be assigned plainclothes inspections ($N = 15{,}284$), while lines with more inspection hours should be assigned uniformed inspections ($N = 811$).

The optimal policy thus recommends plainclothes inspections for the large majority of observations (19{,}369 out of 21{,}100, or 91.8\%). Uniformed inspections are only suggested in two settings: for lines with a low share of foreign residents but a relatively large population, and for lines with a high share of foreign residents and many total inspection hours in the previous month. The role of the share of foreign residents as on of the splitting variables should be interpreted with caution given that this variable may proxy for socioeconomic characteristics more broadly (see Footnote~\ref{fn:foreign}).

\section{Discussion and Conclusion}\label{Discussion}

Fare evasion generates substantial economic losses for public transport companies. To enforce fare payment one core instrument is the organisation of inspection teams. One question which arises is whether inspections are conducted by uniformed (visible) or plainclothes (covert) inspectors. In our study, we assessed the differences between these two strategies on inspection efficiency, defined as the number of detected fare evaders per unit of inspection time.

First, we applied causal machine learning to estimate the effect of the binary treatment of uniformed versus plainclothes inspections on inspection efficiency. The causal machine learning analysis yielded two main findings. First, plainclothes inspections were on average significantly more effective at detecting fare evaders than uniformed inspections, with an estimated ATE of $-$0.156 incidents per hour, corresponding to a relative reduction of approximately 26\% compared to the sample mean. Translating this into monetary terms, the estimated ATE corresponds to approximately 14–16 Swiss francs in additional surcharge revenue per inspection hour under plainclothes inspections, based on the standard first-offence fine of 90–100 Swiss francs \citep[the exact fine amount is set by the operators within the tariff][]{pbg2009}.\footnote{This figure reflects the detection margin and excludes the corresponding fare, which passengers without a valid ticket have to pay on top of the fine.} Second, the heterogeneity analysis when estimating the Best Linear Predictor (BLP) of the conditional average treatment effect (CATE) provides  evidence of systematic variation across contextual characteristics. In addition, GATEs indicate that while the direction of the effect is consistently negative, its magnitude varies across contexts: the most affected group exhibits a GATE of $-0.34$ incidents per hour, compared to $-0.03$ in the least affected group. Overall, these findings suggest that the higher detection efficiency of plainclothes inspections is a robust and pervasive phenomenon across the PostAuto network: while the magnitude of the effect varies across contexts, its direction remains consistent throughout.

Moreover, we applied optimal policy learning to determine the inspection strategy that maximises inspection efficiency across contexts. The results of the policy tree \citep{athey2021policy} indicated that uniformed inspections are only suggested in two settings: for lines with a low share of foreign residents but a relatively large population, and for lines with a high share of foreign residents and many total inspection hours in the previous month.. As these combinations are relatively uncommon in the PostAuto network, the policy tree recommends plainclothes inspections for the large majority of contexts (91.8\%) in order to maximise the number of detected fare evaders per unit of inspection time. However, note that the 91.8\% refers to the share of observations in the analytic dataset rather than to population size or geographic coverage, meaning that contexts recommended for uniformed inspections may nonetheless account for a substantial share of the population served. The split based on past inspection hours can be interpreted as follows: On lines with many inspection hours in the previous month, passengers are more likely to expect an inspection. A uniform therefore deters fewer additional evaders, so the penalty of a uniformed inspection is smaller, which is why the policy tree recommends uniformed inspections. We discuss this deterrence interpretation in more detail below.

When interpreting the results, it is important to note what our outcome measure of inspection efficiency does and does not capture. Inspection efficiency, defined as the number of detected fare evaders per unit of inspection time, quantifies performance at the 'detection margin': conditional on an inspection taking place, how many passengers travelling without a valid ticket are identified per hour of deployed inspection time (including both in-vehicle and between-vehicle time). 

As an alternative interpretation, this difference can also be read as an immediate 'deterrence margin', that is, the extent to which the mere visibility of inspectors induces passengers to purchase or validate a ticket before they are reached by the inspectors. Uniformed inspections may plausibly perform worse at the detection margin precisely because they perform better at this margin: passengers who would otherwise be detected may instead choose to comply, or may adapt their behavior to avoid encountering an inspector altogether \citep{becker1968crime}. Note that the estimate only captures the difference between the deterrence effect of the inspection strategies, as plainclothes inspectors also become identifiable once the inspection has started. This interpretation is only credible if the only difference between the two inspection strategies is the visibility. To test this empirically, we compare the number of inspected passengers per hour by inspection strategy, using a causal forest with the number of inspected passengers per hour as the outcome and inspection strategy as treatment (see Table \ref{tab:ate_throughput} in the Appendix). The result show no significant difference in throughput between the two strategies, which is consistent with the interpretation of the effect as immediate deterrence effect. Note however that this does not include the deterrence effect on future trips, where uniformed inspections might be more effective than plainclothes inspections as well. Our results should therefore be interpreted as evidence on the efficiency of uniformed versus plainclothes inspections in detecting existing fare evaders and on the immediate deterrence effect, rather than as evidence on their effectiveness in reducing the overall incidence of fare evasion over time.

Also when interpreting the results, the external validity of our findings warrants consideration. Our analysis relies exclusively on inspection data from PostAuto, the largest operator of regional bus services in Switzerland. Institutional settings, fare systems, passenger behaviour, inspection procedures, and fare enforcement policies differ considerably across operators and countries, and PostAuto's specific characteristics may not be representative of public transport systems more broadly. Consequently, the relative detection efficiency of uniformed versus plainclothes inspections documented here may not directly generalise to other operators, transport modes (e.g., urban transport or rail), or countries with different institutional and enforcement contexts. Future research should therefore examine whether our findings extend to other operators and settings.

In addition, two limitations of this study warrant mention. First, our findings are conditional on inspections being conducted and therefore do not generalise to contexts in which no inspection activity takes place. Second, the causal identification relies on a selection-on-observables assumption, which (while supported by the institutional context) cannot be directly tested. In particular, potential confounders such as individual inspector characteristics (e.g., experience or ability), one-off local events, and local security conditions are not directly observed in our data, and unobserved confounding from these factors cannot be entirely ruled out. Note that Appendix \ref{Appendix_Robustness} presents a robustness check including lagged inspection efficiency, which (at least partially) addresses concerns about time-invariant line-level confounding potentially related to inspector experience. As the resulting estimates remain close to our main specification, such confounding does not appear to be a major driver of our results.

The main analysis compares uniformed (\textit{Präsenzkontrolle}) versus plainclothes (\textit{Normalkontrolle}) inspections, focusing on the visibility of inspectors as the key dimension of treatment. A complementary perspective concerns not the visibility of inspectors, but whether passengers have the opportunity to evade detection altogether (see Section \ref{background}). In certain inspection formats --- namely coordinated operations such as \textit{Schwerpunktkontrolle}, \textit{Verstärkte Kontrolle}, and \textit{Focus-Security Kontrolle} --- multiple inspectors simultaneously cover all entry and exit points of a vehicle or station, effectively eliminating passengers' ability to avoid inspection. In contrast, standard inspections (\textit{Normalkontrolle} and \textit{Präsenzkontrolle}) are typically conducted by a single inspector or a small team, leaving passengers with the possibility to disembark or move to another carriage before being checked. We therefore define a second treatment variable that captures this distinction: \textit{evasion impossible} (coordinated inspections) versus \textit{evasion possible} (standard inspections), and estimate the causal effect of removing passengers' ability to evade on inspection efficiency.

The estimated average treatment effect of coordinated inspections (evasion impossible) relative to standard inspections (evasion possible) amounts to $-$0.072 incidents per hour (SE = 0.027, $p = 0.008$), suggesting that coordinated inspections detect marginally fewer fare evaders per inspection hour than standard inspections. The 
effect is statistically significant at the 1\% level (see Table \ref{tab:ate_t2}). This finding is not entirely surprising: coordinated operations such as \textit{Schwerpunktkontrolle} or \textit{Verstärkte Kontrolle} typically involve multiple inspectors covering all entry and exit points simultaneously, which implies that total inspection time---including waiting and coordination time across all deployed staff---is substantially higher than for standard inspections. Since inspection efficiency is defined as detected fare evaders per total inspection hour, the denominator is mechanically larger for coordinated operations, which may offset any gains in detection rates. The result should therefore be interpreted with caution and does not necessarily imply that coordinated inspections are less effective at deterring fare evasion overall.

	\newpage
	\bigskip
	
	\bibliographystyle{sageh}
	\bibliography{RogF.bib}
	
	\newpage
	\bigskip

\begin{appendix}
		
		\numberwithin{equation}{section}
		\counterwithin{figure}{section}
		\noindent \textbf{\LARGE Appendices}
	
\section{Additional Figures}\label{Appendix_Figures}

\begin{figure}[H]
	\centering
	\caption{Distribution of Estimated Propensity Scores by Inspection Strategy}\label{fig:pscore}
	\resizebox{0.9\textwidth}{!}{\input{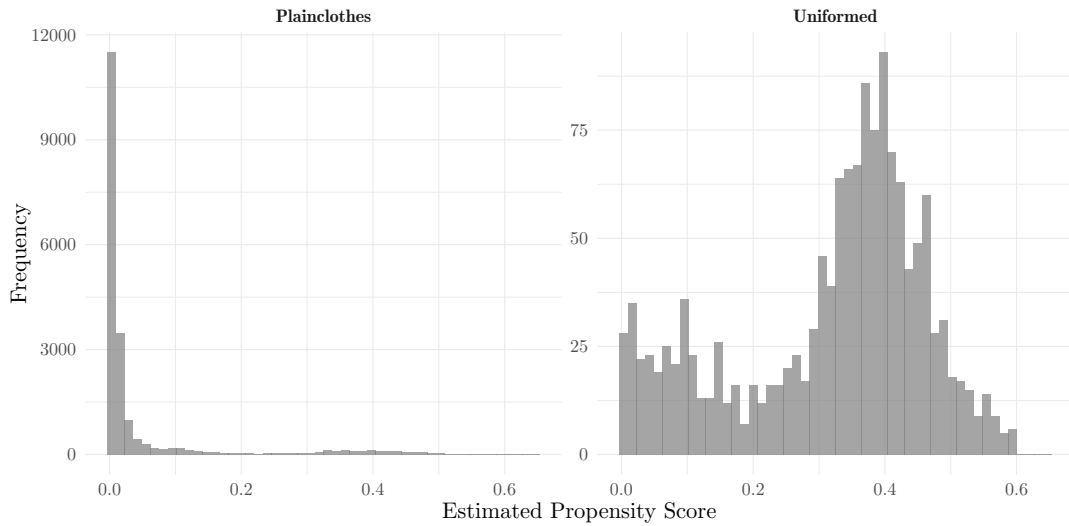}}
	\begin{minipage}{0.9\textwidth}
		\footnotesize\textit{Notes: Propensity scores are estimated via the causal forest. The uniformed group concentrates near zero, reflecting the substantial imbalance in the analytic sample (6.9\% uniformed vs.\ 93.1\% plainclothes). Figure~\ref{fig:pscore_uniformed} provides a rescaled view of the uniformed distribution.}
	\end{minipage}
\end{figure}

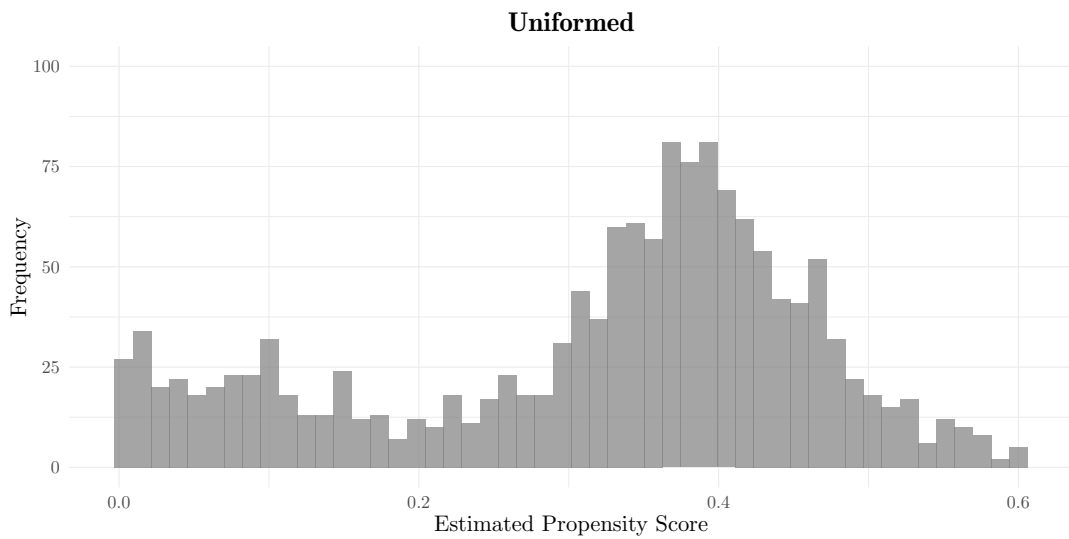
\begin{figure}[H]
	\centering
	\caption{Distribution of Estimated Propensity Scores -- Uniformed Inspections (Rescaled)}\label{fig:pscore_uniformed}
	\resizebox{0.9\textwidth}{!}{
\begin{tikzpicture}[x=1pt,y=1pt]
\definecolor{fillColor}{RGB}{255,255,255}
\path[use as bounding box,fill=fillColor,fill opacity=0.00] (0,0) rectangle (578.16,289.08);
\begin{scope}
\path[clip] (  0.00,  0.00) rectangle (578.16,289.08);
\definecolor{fillColor}{RGB}{255,255,255}

\path[fill=fillColor] (  0.00,  0.00) rectangle (578.16,289.08);
\end{scope}
\begin{scope}
\path[clip] ( 36.11, 30.69) rectangle (572.66,266.40);
\definecolor{drawColor}{gray}{0.92}

\path[draw=drawColor,line width= 0.3pt,line join=round] ( 36.11, 68.19) --
	(572.66, 68.19);

\path[draw=drawColor,line width= 0.3pt,line join=round] ( 36.11,121.76) --
	(572.66,121.76);

\path[draw=drawColor,line width= 0.3pt,line join=round] ( 36.11,175.33) --
	(572.66,175.33);

\path[draw=drawColor,line width= 0.3pt,line join=round] ( 36.11,228.90) --
	(572.66,228.90);

\path[draw=drawColor,line width= 0.3pt,line join=round] (142.87, 30.69) --
	(142.87,266.40);

\path[draw=drawColor,line width= 0.3pt,line join=round] (303.05, 30.69) --
	(303.05,266.40);

\path[draw=drawColor,line width= 0.3pt,line join=round] (463.23, 30.69) --
	(463.23,266.40);

\path[draw=drawColor,line width= 0.6pt,line join=round] ( 36.11, 41.40) --
	(572.66, 41.40);

\path[draw=drawColor,line width= 0.6pt,line join=round] ( 36.11, 94.97) --
	(572.66, 94.97);

\path[draw=drawColor,line width= 0.6pt,line join=round] ( 36.11,148.55) --
	(572.66,148.55);

\path[draw=drawColor,line width= 0.6pt,line join=round] ( 36.11,202.12) --
	(572.66,202.12);

\path[draw=drawColor,line width= 0.6pt,line join=round] ( 36.11,255.69) --
	(572.66,255.69);

\path[draw=drawColor,line width= 0.6pt,line join=round] ( 62.79, 30.69) --
	( 62.79,266.40);

\path[draw=drawColor,line width= 0.6pt,line join=round] (222.96, 30.69) --
	(222.96,266.40);

\path[draw=drawColor,line width= 0.6pt,line join=round] (383.14, 30.69) --
	(383.14,266.40);

\path[draw=drawColor,line width= 0.6pt,line join=round] (543.32, 30.69) --
	(543.32,266.40);
\definecolor{fillColor}{RGB}{127,127,127}

\path[fill=fillColor,fill opacity=0.70] ( 60.50, 41.40) rectangle ( 70.26, 99.26);

\path[fill=fillColor,fill opacity=0.70] ( 70.26, 41.40) rectangle ( 80.01,114.26);

\path[fill=fillColor,fill opacity=0.70] ( 80.01, 41.40) rectangle ( 89.77, 84.26);

\path[fill=fillColor,fill opacity=0.70] ( 89.77, 41.40) rectangle ( 99.52, 88.54);

\path[fill=fillColor,fill opacity=0.70] ( 99.52, 41.40) rectangle (109.28, 79.97);

\path[fill=fillColor,fill opacity=0.70] (109.28, 41.40) rectangle (119.03, 84.26);

\path[fill=fillColor,fill opacity=0.70] (119.03, 41.40) rectangle (128.79, 90.69);

\path[fill=fillColor,fill opacity=0.70] (128.79, 41.40) rectangle (138.54, 90.69);

\path[fill=fillColor,fill opacity=0.70] (138.54, 41.40) rectangle (148.30,109.97);

\path[fill=fillColor,fill opacity=0.70] (148.30, 41.40) rectangle (158.05, 79.97);

\path[fill=fillColor,fill opacity=0.70] (158.05, 41.40) rectangle (167.81, 69.26);

\path[fill=fillColor,fill opacity=0.70] (167.81, 41.40) rectangle (177.56, 69.26);

\path[fill=fillColor,fill opacity=0.70] (177.56, 41.40) rectangle (187.32, 92.83);

\path[fill=fillColor,fill opacity=0.70] (187.32, 41.40) rectangle (197.08, 67.12);

\path[fill=fillColor,fill opacity=0.70] (197.08, 41.40) rectangle (206.83, 69.26);

\path[fill=fillColor,fill opacity=0.70] (206.83, 41.40) rectangle (216.59, 56.40);

\path[fill=fillColor,fill opacity=0.70] (216.59, 41.40) rectangle (226.34, 67.12);

\path[fill=fillColor,fill opacity=0.70] (226.34, 41.40) rectangle (236.10, 62.83);

\path[fill=fillColor,fill opacity=0.70] (236.10, 41.40) rectangle (245.85, 79.97);

\path[fill=fillColor,fill opacity=0.70] (245.85, 41.40) rectangle (255.61, 64.97);

\path[fill=fillColor,fill opacity=0.70] (255.61, 41.40) rectangle (265.36, 77.83);

\path[fill=fillColor,fill opacity=0.70] (265.36, 41.40) rectangle (275.12, 90.69);

\path[fill=fillColor,fill opacity=0.70] (275.12, 41.40) rectangle (284.87, 79.97);

\path[fill=fillColor,fill opacity=0.70] (284.87, 41.40) rectangle (294.63, 79.97);

\path[fill=fillColor,fill opacity=0.70] (294.63, 41.40) rectangle (304.39,107.83);

\path[fill=fillColor,fill opacity=0.70] (304.39, 41.40) rectangle (314.14,135.69);

\path[fill=fillColor,fill opacity=0.70] (314.14, 41.40) rectangle (323.90,120.69);

\path[fill=fillColor,fill opacity=0.70] (323.90, 41.40) rectangle (333.65,169.97);

\path[fill=fillColor,fill opacity=0.70] (333.65, 41.40) rectangle (343.41,172.12);

\path[fill=fillColor,fill opacity=0.70] (343.41, 41.40) rectangle (353.16,163.55);

\path[fill=fillColor,fill opacity=0.70] (353.16, 41.40) rectangle (362.92,214.97);

\path[fill=fillColor,fill opacity=0.70] (362.92, 41.40) rectangle (372.67,204.26);

\path[fill=fillColor,fill opacity=0.70] (372.67, 41.40) rectangle (382.43,214.97);

\path[fill=fillColor,fill opacity=0.70] (382.43, 41.40) rectangle (392.18,189.26);

\path[fill=fillColor,fill opacity=0.70] (392.18, 41.40) rectangle (401.94,174.26);

\path[fill=fillColor,fill opacity=0.70] (401.94, 41.40) rectangle (411.70,157.12);

\path[fill=fillColor,fill opacity=0.70] (411.70, 41.40) rectangle (421.45,131.40);

\path[fill=fillColor,fill opacity=0.70] (421.45, 41.40) rectangle (431.21,129.26);

\path[fill=fillColor,fill opacity=0.70] (431.21, 41.40) rectangle (440.96,152.83);

\path[fill=fillColor,fill opacity=0.70] (440.96, 41.40) rectangle (450.72,109.97);

\path[fill=fillColor,fill opacity=0.70] (450.72, 41.40) rectangle (460.47, 88.54);

\path[fill=fillColor,fill opacity=0.70] (460.47, 41.40) rectangle (470.23, 79.97);

\path[fill=fillColor,fill opacity=0.70] (470.23, 41.40) rectangle (479.98, 73.54);

\path[fill=fillColor,fill opacity=0.70] (479.98, 41.40) rectangle (489.74, 77.83);

\path[fill=fillColor,fill opacity=0.70] (489.74, 41.40) rectangle (499.49, 54.26);

\path[fill=fillColor,fill opacity=0.70] (499.49, 41.40) rectangle (509.25, 67.12);

\path[fill=fillColor,fill opacity=0.70] (509.25, 41.40) rectangle (519.01, 62.83);

\path[fill=fillColor,fill opacity=0.70] (519.01, 41.40) rectangle (528.76, 58.54);

\path[fill=fillColor,fill opacity=0.70] (528.76, 41.40) rectangle (538.52, 45.69);

\path[fill=fillColor,fill opacity=0.70] (538.52, 41.40) rectangle (548.27, 52.11);
\end{scope}
\begin{scope}
\path[clip] (  0.00,  0.00) rectangle (578.16,289.08);
\definecolor{drawColor}{gray}{0.30}

\node[text=drawColor,anchor=base east,inner sep=0pt, outer sep=0pt, scale=  0.88] at ( 31.16, 38.37) {0};

\node[text=drawColor,anchor=base east,inner sep=0pt, outer sep=0pt, scale=  0.88] at ( 31.16, 91.94) {25};

\node[text=drawColor,anchor=base east,inner sep=0pt, outer sep=0pt, scale=  0.88] at ( 31.16,145.51) {50};

\node[text=drawColor,anchor=base east,inner sep=0pt, outer sep=0pt, scale=  0.88] at ( 31.16,199.09) {75};

\node[text=drawColor,anchor=base east,inner sep=0pt, outer sep=0pt, scale=  0.88] at ( 31.16,252.66) {100};
\end{scope}
\begin{scope}
\path[clip] (  0.00,  0.00) rectangle (578.16,289.08);
\definecolor{drawColor}{gray}{0.30}

\node[text=drawColor,anchor=base,inner sep=0pt, outer sep=0pt, scale=  0.88] at ( 62.79, 19.68) {0.0};

\node[text=drawColor,anchor=base,inner sep=0pt, outer sep=0pt, scale=  0.88] at (222.96, 19.68) {0.2};

\node[text=drawColor,anchor=base,inner sep=0pt, outer sep=0pt, scale=  0.88] at (383.14, 19.68) {0.4};

\node[text=drawColor,anchor=base,inner sep=0pt, outer sep=0pt, scale=  0.88] at (543.32, 19.68) {0.6};
\end{scope}
\begin{scope}
\path[clip] (  0.00,  0.00) rectangle (578.16,289.08);
\definecolor{drawColor}{RGB}{0,0,0}

\node[text=drawColor,anchor=base,inner sep=0pt, outer sep=0pt, scale=  1.10] at (304.39,  7.64) {Estimated Propensity Score};
\end{scope}
\begin{scope}
\path[clip] (  0.00,  0.00) rectangle (578.16,289.08);
\definecolor{drawColor}{RGB}{0,0,0}

\node[text=drawColor,rotate= 90.00,anchor=base,inner sep=0pt, outer sep=0pt, scale=  1.10] at ( 13.08,148.55) {Frequency};
\end{scope}
\begin{scope}
\path[clip] (  0.00,  0.00) rectangle (578.16,289.08);
\definecolor{drawColor}{RGB}{0,0,0}

\node[text=drawColor,anchor=base,inner sep=0pt, outer sep=0pt, scale=  1.32] at (304.39,274.47) {\bfseries Uniformed};
\end{scope}
\end{tikzpicture}}
	\begin{minipage}{0.9\textwidth}
		\footnotesize\textit{Notes: Y-axis truncated at 100 to improve visibility.}
	\end{minipage}
\end{figure}

\section{Additional Tables}\label{Appendix_Tabels}

\begin{table}[H]
	\centering
	\caption{Average Treatment Effect -- Treated Target Sample (Robustness)}
	\label{tab:ate_robustness}
	\begin{tabular}{lccc}
		\hline
		& Estimate & Std. Error & $p$-value \\
		\hline
		Causal Forest (treated) & $-$0.152 & 0.029 & $<$0.001 \\
		\hline
		Number of observations & \multicolumn{3}{c}{21,100} \\
		\hline
	\end{tabular}
	\begin{minipage}{0.7\textwidth}
		\footnotesize\textit{Notes: The average treatment effect is estimated using the causal forest with target sample set to the treated population ($D = 1$ = uniformed). The $p$-value is based on a two-sided $z$-test.}
	\end{minipage}
\end{table}

\begin{table}[H]
	\centering
	\caption{Average Treatment Effect: Evasion Impossible versus Evasion Possible (T2)}
	\label{tab:ate_t2}
	\begin{tabular}{lccc}
		\hline
		& Estimate & Std. Error & $p$-value \\
		\hline
		Causal Forest (overlap) & $-$0.072 & 0.027 & 0.008 \\
		Causal Forest (treated) & $-$0.075 & 0.025 & 0.003 \\
		\hline
	\end{tabular}
	\begin{minipage}{0.75\textwidth}
		\footnotesize\textit{Notes: The average treatment effect measures the effect of coordinated inspections (evasion impossible: \textit{Schwerpunktkontrolle}, \textit{Verstärkte Kontrolle}, \textit{Focus-Security Kontrolle}) relative to standard inspections (evasion possible: \textit{Normalkontrolle}, \textit{Präsenzkontrolle}). $D = 1$ = evasion impossible, $D = 0$ = evasion possible. The $p$-value is based on a two-sided $z$-test.}
	\end{minipage}
\end{table}
	
\begin{table}[H]
	\centering
	\caption{Average Treatment Effect on Inspection Throughput}
	\label{tab:ate_throughput}
	\begin{tabular}{lccc}
		\hline
		& Estimate & Std. Error & $p$-value \\
		\hline
		Causal Forest (overlap) & $-$0.724 & 0.612 & 0.237 \\
		Causal Forest (treated) & $-$0.597 & 0.625 & 0.340 \\
		\hline
	\end{tabular}
	\begin{minipage}{0.75\textwidth}
		\footnotesize\textit{Notes: The outcome is the number of inspected passengers per hour of inspection time, including time spent transferring between vehicles. The average treatment effect measures the effect of uniformed inspections (\textit{Pr\"{a}senzkontrolle}, $D = 1$) relative to plainclothes inspections (\textit{Normalkontrolle}, $D = 0$). The $p$-value is based on a two-sided $z$-test.}
	\end{minipage}
\end{table}

\begin{table}[H]
	\centering
	\small
	\caption{Covariates used in the causal forest}
	\label{tab:covariates}
	\begin{tabular*}{\textwidth}{@{\extracolsep{\fill}}llc@{}}
		\hline
		Variable & Summary statistics & No. \\
		\hline
		\textit{Time of day} & & \\
		Time slot (morning commute to night) & Indicator & 7 \\
		\hline
		\textit{Sociodemographic characteristics} & & \\
		Population & Six statistics & 6 \\
		Share of foreign population & Six statistics & 6 \\
		Youth dependency ratio & Six statistics & 6 \\
		Social assistance rate & Six statistics & 6 \\
		Unemployment rate & Six statistics & 6 \\
		\hline
		\textit{Public transport characteristics} & & \\
		GA ownership rate & Six statistics & 6 \\
		Half-fare card ownership rate & Six statistics & 6 \\
		Regional season ticket ownership rate & Six statistics & 6 \\
		PT accessibility (classes A--D and none) & Share of stops & 5 \\
		\hline
		\textit{Spatial and cultural typology} & & \\
		Degree of urbanisation (urban, intermediate, rural) & Share of stops & 3 \\
		Language region (German, French, Italian, Romansh) & Share of stops & 4 \\
		Touristic classification (touristic, non-touristic) & Share of stops & 2 \\
		\hline
		\textit{Prior inspection activity (previous month)} & & \\
		Inspection hours, total & Sum & 1 \\
		Inspection hours, same time slot & Sum & 1 \\
		\hline
		\textit{Prior inspection efficiency (previous month)}$^{\dagger}$ & & \\
		Inspection efficiency, total$^{\dagger}$ & Ratio & 1 \\
		Inspection efficiency, same time slot$^{\dagger}$ & Ratio & 1 \\
		\hline
		Total (main specification) & & 71 \\
		Total (robustness check) & & 73 \\
		\hline
	\end{tabular*}
	\begin{tablenotes}
		\small
		\item Notes: All covariates are measured at the line level. Six statistics denotes mean, median, minimum, maximum and the first and third quartile across the stops of a line. Categorical attributes enter as the share of a line's stops in each category. Inspection efficiency is measured as incidents per hour. $^{\dagger}$ Included only in the robustness check in Table \ref{tab:ate_lag_efficiency}.
	\end{tablenotes}
\end{table}

\section{Robustness Check - Controlling for Prior Inspection Efficiency}\label{Appendix_Robustness}

Operators and inspectors may inspect lines with high previous inspection efficiency more often. Therefore, as a robustness check, we include inspection efficiency in the previous month, both across all time slots and restricted to the same time slot as the current observation. This is not our main specification because not every line is inspected in every month and every time slot. The resulting missing values reduce the sample from 21,100 to 14,430 observations. The observations that drop out are most likely from lines with sparse inspection coverage.

The overlap-sample estimate from this robustness check ($-$0.179, Table \ref{tab:ate_lag_efficiency}) is somewhat larger in absolute terms than the corresponding estimate from our main specification ($-$0.156). However, when running the main specification on the reduced sample (due to missing lagged efficiency values), the resulting estimate ($-$0.178, Table \ref{tab:ate_reduced_sample}) is almost identical to the one including lagged inspection efficiency, indicating that the larger effect in absolute terms is most likely due to the change in sample composition and not driven by the inclusion of previous inspection efficiency.

\begin{table}[H]
	\centering
	\caption{Average Treatment Effect - Including Lagged Inspection Efficiency}
	\label{tab:ate_lag_efficiency}
	\begin{tabular}{lccc}
		\hline
		& Estimate & Std. Error & $p$-value \\
		\hline
		Causal Forest (overlap) & $-$0.179 & 0.035 & $<$0.001 \\
		Causal Forest (treated) & $-$0.184 & 0.034 & $<$0.001 \\
		\hline
		Number of observations& \multicolumn{3}{c}{14,430} \\
		\hline
	\end{tabular}
	\begin{minipage}{0.75\textwidth}
		\footnotesize\textit{Notes: The average treatment effect is estimated using the causal forest. The $p$-value is based on a two-sided $z$-test.}
	\end{minipage}
\end{table}

\begin{table}[H]
	\centering
	\caption{Average Treatment Effect - Reduced Sample}
	\label{tab:ate_reduced_sample}
	\begin{tabular}{lccc}
		\hline
		& Estimate & Std. Error & $p$-value \\
		\hline
		Causal Forest (overlap) & $-$0.178 & 0.034 & $<$0.001 \\
		Causal Forest (treated) & $-$0.180 & 0.035 & $<$0.001 \\
		\hline
		Number of observations  & \multicolumn{3}{c}{14,430} \\
		\hline
	\end{tabular}
	\begin{minipage}{0.75\textwidth}
		\footnotesize\textit{Notes: The average treatment effect is estimated using the causal forest. The $p$-value is based on a two-sided $z$-test.}
	\end{minipage}
\end{table}

\end{appendix}
\end{document}